\documentclass[titlepage,12pt]{utarticle}
\usepackage[all]{xy}
\usepackage[]{hyperref} 

\numberwithin{equation}{section}
\begin{document}
\preprint{
  UTTG--18--06\\
  {\tt hep-th/0701244}\\
}
\title{
  Heterotic Compactifications with Principal Bundles\vphantom{y}\\
  for General Groups and General Levels
}
\author{
  Jacques Distler
    \thanks{Work supported in part by NSF Grant PHY-0455649.}
    \address{
      Theory Group, Physics Department\\
      University of Texas at Austin\\
      Austin, TX 78712 USA\\
      {~}\\
      \email{distler@golem.ph.utexas.edu}
    }
    and Eric Sharpe
    \address{
      Departments of Physics, Mathematics\\
      University of Utah\\
      Salt Lake City, UT  84112\\
      {~}\\
      \email{ersharpe@math.utah.edu}
    }
}
\date{January 28, 2007}

\Abstract{
We examine to what extent heterotic string worldsheets
can describe arbitrary $E_8 \times E_8$ gauge fields.
The traditional construction of heterotic strings builds each
$E_8$ via a $\mathrm{Spin}(16)/\BZ_2$ subgroup, typically realized
as a current algebra by left-moving fermions, and as a result, only $E_8$ gauge fields reducible to $\mathrm{Spin}(16)/\BZ_2$ gauge
fields are directly realizable in standard constructions.
However, there exist perturbatively consistent $E_8$ gauge fields which can not be reduced to $\mathrm{Spin}(16)/\BZ_2$,
and so cannot be described within standard heterotic worldsheet constructions. 
A natural question to then ask is whether there exists any (0,2) SCFT 
that can describe such $E_8$ gauge fields. 
To answer this question, we first show how each
ten-dimensional $E_8$ partition function can be built up using other 
subgroups than $\mathrm{Spin}(16)/\BZ_2$, then construct
``fibered WZW models'' which allow us to explicitly couple current algebras for general groups and general levels to heterotic strings. 
This technology gives us a very general approach to handling
heterotic compactifications with arbitrary principal bundles.
It also gives us a physical realization of some elliptic genera 
constructed recently by Ando and Liu. 
}

\maketitle


\tableofcontents

\newpage

\section{Introduction}\label{sec:intro}
In the last few years there has been a great deal of interest in 
the `landscape' program as a mechanism for extracting phenomenological
predictions from string theory by doing statistics on sets of potential
vacua.  One of the potential problems with this program is that the
potential vacua are classified by low-energy effective supergravity
theories, and it is not clear to what extent all possible supergravity
theories can be described within string theory \cite{bankstalk,vafaswamp}.

In this paper we will analyze examples potentially lacking
UV-completions, in heterotic strings.
Specifically, we begin by observing that not all principal $E_8$
bundles with connection that satisfy the conditions for a supergravity
vacuum can be described within traditional formulations of perturbative
heterotic string theory.  The basic problem is that traditional heterotic
string constructions build each $E_8$ from a $\mathrm{Spin}(16)/\BZ_2$
subgroup, and so can only describe those $E_8$ bundles with connection
reducible to $\mathrm{Spin}(16)/\BZ_2$.  However, not all 
principal $E_8$ bundles with connection are reducible to 
$\mathrm{Spin}(16)/\BZ_2$, and those which cannot be so reduced,
cannot be described within traditional heterotic string constructions.

That lack of reducibility suggests there may be a problem with
the existence of UV completions for such heterotic supergravity theories.
However, we point out that there exists evidence from string duality
that suggests UV completions for these exotic heterotic supergravities
should still exist, and in the rest of the paper we go on to build
new worldsheet theories which can be used to describe more general
$E_8$ bundles with connection than the traditional constructions.

This paper can be broken into three main sections:
\begin{enumerate}
\item After initially reviewing the construction of $E_8$ bundles
via $\mathrm{Spin}(16)/\BZ_2$ subbundles in section~\ref{wsobs},
in sections~\ref{ppale8} and \ref{e8conn} we analyze the extent to
which $E_8$ bundles with connection can be described by the usual
fermionic realization of the heterotic string.  We find that there
is a topological obstruction to describing certain $E_8$ bundles
in dimension 10, but more alarmingly, in lower dimensions there is
an obstruction to describing all gauge fields.
In particular, we describe some examples of $E_8$ bundles with connection
in dimension less than 10 which satisfy the usual constraints for
a perturbative string vacuum but which cannot be described by 
traditional 
worldsheet realizations of the heterotic string.
This seems to suggest that not all $E_8$ bundles with connection
can be realized perturbatively.
However, in section~\ref{fthydual} we observe that other evidence such as
F theory calculations suggests that, in fact, the other
$E_8$ bundles with connection {\it can} be realized perturbatively,
just not with traditional constructions.  In the rest of the paper
we describe alternative constructions of heterotic strings which
can be used to describe the `exceptional' gauge fields above.
\item The next part of this paper, section~\ref{alt10d},
is a discussion of alternative constructions of each $E_8$ in
a ten-dimensional theory.  The usual fermionic construction builds
each $E_8$ using a $\mathrm{Spin}(16)/\BZ_2$ subgroup -- the left-moving
worldsheet fermions realize a $\mathrm{Spin}(16)$ and a left-moving
$\BZ_2$ orbifold realizes the $/\BZ_2$.
However, there are other subgroups of $E_8$ that can also be used
instead, such as $( SU(5) \times SU(5) )/\BZ_5$ and
$SU(9)/\BZ_3$.  
At the level of characters of affine algebras,
such constructions have previously been described in {\it e.g.}
\cite{kacsan}.
We check that the ten-dimensional partition function
of current algebras realizing other\footnote{For example,
the $(SU(5) \times SU(5))/\BZ_5$ subgroup describes a 
$\BZ_5$ orbifold of an $SU(5) \times SU(5)$ current algebra,
just as the traditional $\mathrm{Spin}(16)/\BZ_2$ subgroup describes
a $\BZ_2$ orbifold of a $\mathrm{Spin}(16)$ current algebra
(realized by free fermions).} 
$E_8$ subgroups correctly
reproduces the usual self-dual modular invariant
partition function.
\item To make this useful we need to understand how more general
current algebras can be fibered nontrivially over a base,
and so in the third part of this paper, sections~\ref{symmfibwzw}
and \ref{chirfibwzw}, we develop\footnote{After the initial publication
of this paper it was pointed out to us that chiral fibered WZW models
with $(0,1)$ supersymmetry have been previously considered,
under the name ``lefton, righton Thirring models,'' see for example
\cite{gates1,gates2,gates3,gates4,gates5}.  We develop the notion
further, by studying anomaly cancellation, spectra, elliptic genera,
and so forth in chiral fibered WZW models with $(0,2)$ supersymmetry.} 
and analyze
``fibered WZW models,'' which allow us to work with heterotic $(0,2)$
supersymmetric SCFT's in which the left-movers couple to
some general $G$-current algebra at level $k$, for general $G$ and $k$,
fibered nontrivially over the target space.
Only for certain $G$ and $k$ can these CFT's be used in critical heterotic
string compactifications, but the general result is of interest to
the study of heterotic CFT's.  The construction of these theories
is interesting:  bosonizing the left-movers into a WZW model
turns quantum features of fermionic realizations into classical
features, and so to understand the resulting theory requires mixing
such classical features against quantum effects such as the chiral
anomaly of the right-moving fermions. 
This construction also gives us a physical realization of some
elliptic genera constructed in the mathematics community previously.
The generalization of the anomaly cancellation condition that we
derive in our model, for example, was independently derived by
mathematicians thinking about generalizations of elliptic genera.
\end{enumerate}

To a large extent, the three parts of this paper can nearly be read
independently of one another.  For example, readers who only wish to
learn about fibered WZW  model constructions should be able to
read sections~\ref{symmfibwzw} and \ref{chirfibwzw} without having
mastered the earlier material.

Higher-level Kac-Moody algebras in heterotic
compactifications have been considered previously in the
context of free fermion models, see for example \cite{lewellen,dienes} which
discuss their phenomenological virtues.  In \cite{dienes}, for example,
the higher-level Kac-Moody algebras are constructed by starting
with critical heterotic strings realized in the usual fashion
and then orbifolding in such a way as to realize higher-level
Kac-Moody algebras from within the original level one structure.
However, in each of those previous works the higher-level Kac-Moody
algebras were all essentially embedded in an ambient level one algebra,
the ordinary $E_8$ algebra.
We are not aware of any previous work discussing heterotic compactifications
with higher-level Kac-Moody algebras that realize those algebras
directly, without an embedding into some ambient algebra,
as we do in this paper with `fibered WZW' models.

\section{Worldsheet obstruction in standard constructions}\label{wsobs}
How does one describe an $E_8$ bundle on the worldsheet?
It is well-known how to construct the $E_8$ current algebra,
and bundles with structure groups of the form $SU(n) \times U(1)^m$ are
also understood in this language, but to understand more exotic cases,
let us carefully work through the details for general nontrivial bundles.

For each $E_8$, there are\footnote{There is, of course, also a representation
of the bosonic string in terms of chiral abelian bosons.
However, that abelian bosonic representation can describe even fewer
bundles with connection than the fermionic representation -- essentially,
only those in which the bundle with connection is reducible to a maximal
torus -- and so we shall focus on the fermionic presentation.} 
16 left-moving fermions which couple to the
pullback of a real vector bundle on the target space associated
to a principal $\mathrm{Spin}(16)$ bundle.
The worldsheet left-moving fermion kinetic terms have the form
\begin{equation*}
h_{\alpha \beta} \lambda_-^{\alpha} D \lambda_-^{\beta}
\end{equation*}
where $h_{\alpha \beta}$ is a fiber metric on a real rank 16 vector bundle,
and $D$ is a covariant derivative which implicitly includes the pullback
of a connection on such a bundle, so we see that we can describe
only $\mathrm{Spin}(16)$ gauge fields.  
The worldsheet GSO projection is equivalent to a $\BZ_2$ orbifold
in which each of those fermions is acted upon by a sign.
Performing the GSO projection is therefore equivalent to projecting
the $\mathrm{Spin}(16)$ bundle to a $\mathrm{Spin}(16)/\BZ_2$ bundle,
and the surviving adjoint and spinor representations of 
$\mathrm{Spin}(16)/\BZ_2$ are built into an $E_8$ bundle,
into which the $\mathrm{Spin}(16)/\BZ_2$ bundle
injects.
(The $\mathrm{Spin}(32)/\BZ_2$ heterotic string is much simpler;
the 32 left-moving spinors couple to a vector bundle associated
to a principal $\mathrm{Spin}(32)$ bundle, and the GSO projection
projects to $\mathrm{Spin}(32)/\BZ_2$.)

Factors of $\BZ_2$ will play an important role in what follows,
so let us take a moment to carefully check the statement above.
Of the groups $O(16)$, $SO(16)$, $\mathrm{Spin}(16)$,
and $\mathrm{Spin}(16)/\BZ_2$, only $\mathrm{Spin}(16)/\BZ_2$
is a subgroup of $E_8$ \cite{bryantpriv,adams}, 
so after performing the GSO projection
we had better recover a $\mathrm{Spin}(16)/\BZ_2$ bundle.
Also, the fact that the adjoint representation of $E_8$ decomposes
into the adjoint representation of $so(16)$ plus one chiral spinor
gives us another clue -- if the subgroup were $SO(16)$, then no spinors
could appear in the decomposition.  The $\BZ_2$ quotient in
$\mathrm{Spin}(16)/\BZ_2$ projects out one of the chiral spinors
but not the other, giving us precisely the matter that we see
perturbatively.
Furthermore, $\mathrm{Spin}(16)/\BZ_2$ does not have a 
16-dimensional representation, so the left-moving fermions cannot
be in a vector bundle associated to a principal $\mathrm{Spin}(16)/\BZ_2$
bundle.  Instead, they couple to a $\mathrm{Spin}(16)$ bundle,
and the GSO projection plays a crucial role.

Any data about a bundle with connection on the target space must be
encoded in the fermion kinetic terms
\begin{equation*}
h_{\alpha \beta} \lambda_-^{\alpha} D \lambda_-^{\beta}
\end{equation*}
Since the only data encoded concerns $\mathrm{Spin}(16)$ bundles,
if we had an $E_8$ bundle with connection that could not be reduced
to $\mathrm{Spin}(16)/\BZ_2$ and then lifted to $\mathrm{Spin}(16)$,
we would not be able to describe it on the worldsheet using the
conventional fermionic realization of the heterotic string.

So far we have described what worldsheet structures define the
$E_8$ bundle on the target space.
Let us now think about the reverse operation.
Given an $E_8$ bundle, what does one do to construct the corresponding
heterotic string?
First, one reduces the structure group from $E_8$ to
$\mathrm{Spin}(16)/\BZ_2$, if possible, and then lifts from
$\mathrm{Spin}(16)/\BZ_2$ to $\mathrm{Spin}(16)$, if possible.
The resulting $\mathrm{Spin}(16)$ bundle defines the left-moving 
worldsheet fermions.

The catch is that not all $E_8$ bundles are reducible to 
$\mathrm{Spin}(16)/\BZ_2$, and not all
$\mathrm{Spin}(16)/\BZ_2$ bundles can be lifted to
$\mathrm{Spin}(16)$ bundles.  The second obstruction is defined by
an analogue of a Stiefel-Whitney class, which is more or less reasonably
well understood.  We will be primarily concerned in this paper with the first
obstruction, which to our knowledge has not been discussed in the physics
literature previously.

\section{Principal $E_8$ bundles}\label{ppale8}

\subsection{Reducibility of principal $E_8$ bundles}
In this section we shall briefly outline\footnote{We are indebted to
A.~Henriques for a lengthy discussion in which he explained the
points of this section, and for giving us 
permission to repeat  
his homotopy analysis here.} 
the technical issues involved
in computing the obstruction to reducing an $E_8$ bundle to 
a $\mathrm{Spin}(16)/\BZ_2$ bundle.  We shall find that the only
obstruction is an element of $H^{10}(M, \BZ_2)$, where $M$ is the
spacetime ten-manifold on which the $E_8$ bundle lives.

An $E_8$ bundle is the same thing as a map $M \rightarrow BE_8$.
In order to reduce the structure group of the bundle to $\mathrm{Spin}(16)/\BZ_2$, we want to lift the map above to a map
$M \rightarrow B\mathrm{Spin}(16)/\BZ_2$.  In
fact, for our purposes, we can equivalently consider $B SO(16)$,
which is technically somewhat simpler.

In general, if $M$ is simply-connected (which we shall assume
throughout this section), then the obstructions to reducing a principal
$G$-bundle on $M$ to a principal $H$-bundle for $H \subset G$ live in
$H^k(M, \pi_{k-1}(G/H))$, which can be proven with
Postnikov towers.  Since this technology is not widely
used in the physics community, let us expound upon this method
for $H=1$, and study 
the obstructions to trivializing a principal $G$ bundle which,
from the general statement above, live in
$H^k(M, \pi_{k-1}(G))$.  It is well-known that a principal $G$ bundle can
be trivialized if its characteristic classes vanish, and so one would
be tempted to believe that the group $H^k(M, \pi_{k-1}(G))$ correspond
to characteristic classes, but the correct relationship\footnote{We would
like to thank M.~Ando for a patient explanation of this point.}
is more complicated.  In the case of $E_8$ bundles and
$U(n)$ bundles, it is straightforward to check that the groups in which the
obstructions live are the same as the ones the characteristic classes
live in, making the distinction obscure:
for $E_8$, since $\pi_3(E_8) = \BZ$ is
the only nonzero homotopy group in dimension ten or less,
the obstructions to trivialing a principal $E_8$ bundle on a manifold
of dimension ten or less live in $H^4(M, \BZ)$, same as the characteristic
class,
and, for $U(n)$ bundles,
$\pi_i(U(n))$ is $\BZ$ for $i$ odd and less than $2n$,
so the obstructions to trivializing $U(n)$ bundles live in $H^{even}(M,
\BZ)$, the same groups as the Chern classes.
Principal $O(n)$ bundles are more confusing, and better illustrate
the distinction between obstructions and characteristic classes.  The homotopy
groups 
\begin{equation*}
\pi_{3 + 8k}(O(n)) \: = \: \BZ \: = \: \pi_{7+8k}(O(n))
\end{equation*}
(for $n$ sufficiently large)
and the corresponding obstructions correspond to the Pontryagin classes
in degrees any multiple of four.  However, there are additional 
$\BZ_2$-valued characteristic classes of $O(n)$ bundles,
known as the Stiefel-Whitney classes, and 
\begin{equation*}
\pi_{0+8k}(O(n)) \: = \: \BZ_2 \: = \: \pi_{1+8k}(O(n))
\end{equation*}
(for $n$ sufficiently large)
corresponding to the first two Stiefel-Whitney classes $w_1$, $w_2$.
However, other homotopy groups vanish
\begin{equation*}
\pi_{2+8k}(O(n)) \: = \: 0 \: = \: \pi_{4+8k}(O(n)) \: = \:
\pi_{5+8k}(O(n))
\end{equation*}
and so there are no obstructions living in $H^3(M,\BZ_2)$,
$H^5(M,\BZ_2)$, or $H^6(M, \BZ_2)$, for example,
despite the fact that
there are Stiefel-Whitney classes in those degrees.
An $O(n)$ bundle can be trivialized only if its characteristic classes
all vanish, and yet we have found no obstructions corresponding to
many Stiefel-Whitney classes, which appears to be a contradiction.
Part of the resolution is that the relationship between characteristic
classes and obstructions is complicated:  for example, the degree four
obstruction is $p_1/2$, and is only defined if the lower-order
obstructions vanish (so that $p_1$ is even).
Higher-order obstructions have an even more
complicated relationship.  At the same time, one can use Steenrod square
operations and the Wu formula to determine many higher-order Stiefel-Whitney
classes from lower ones -- for example, if $w_1=w_2=0$ then necessarily
$w_3=0$.
The upshot of all this is that if the obstructions all vanish,
then the characteristic classes will all vanish, 
and so the bundle is trivializable, and there is no contradiction.

In any event, the obstructions to reducing a principal $E_8$ bundle to
a principal $\mathrm{Spin}(16)/\BZ_2$ bundle live
in $H^k(M,\pi_{k-1}(F))$, where 
$F = E_8/(\mathrm{Spin}(16)/\BZ_2)$ denotes the
fiber of 
\begin{equation*}
B \, \mathrm{Spin}(16)/\BZ_2 \: \longrightarrow \: BE_8. 
\end{equation*}
We can compute the homotopy groups of that quotient using
the long exact
sequence in homotopy induced by the fiber
sequence 
\begin{equation*}
E_8/(\mathrm{Spin}(16)/\BZ_2) \: \longrightarrow \: 
B\mathrm{Spin}(16)/\BZ_2 \: \longrightarrow \: BE_8.
\end{equation*}
                                                                                
One can compute the following:
\begin{center} 
\begin{tabular}{ccccccccccccc}                                                
$\pi_i$ for $i=$ & 1&  2&  3&  4&  5&  6&  7&  8&  9&  10&  11&  12 \\ \hline
$E_8/(\mathrm{Spin}(16)/\BZ_2)$ & 0 & $\BZ_2$ & 0&  0&  0&  0&  0&  $\BZ$ &  $\BZ_2$ &  $\BZ_2$ & 0 &  $\BZ$ \\
$B\mathrm{Spin}(16)/\BZ_2$ & 0 & $\BZ_2$ &  0 & $\BZ$ &  0 & 0 & 0 &
$\BZ$ & $\BZ_2$ & $\BZ_2$ &  0 &  $\BZ$ \\
$BE_8$ &   0 & 0 & 0 & $\BZ$ &  0 &  0&  0&  0 & 0 &  0 &  0 &  0 \\
\end{tabular}                                                                  
\end{center}

We used the following facts to compute this table.
                                                                                
First, we know that $E_8$ looks like a $K(\BZ,3)$ up to dimension 14,
and we also know $\pi_*(BSO)$ by Bott periodicity (see for example
\cite{hatcher}[section 4.2]).  So, to
determine the long exact senquence in the relevant range, we only need to
compute $\pi_4(B\mathrm{Spin}(16)/\BZ_2) \rightarrow \pi_4(BE_8)$.
                                                                                
It turns out that $\pi_4(B\mathrm{Spin}(16)/\BZ_2) \rightarrow \pi_4(BE_8)$ 
is an isomorphism.  This is the
case since $\mathrm{Spin}(16)/\BZ_2 \rightarrow E_8$ comes from an 
inclusion of simply laced root systems
and the $SU(2)$s coming from the roots are the generators of $\pi_3$.

The obstructions in $H^k(M,\pi_{k-1}(F))$ are pulled
back from universal obstructions 
\begin{equation*}
H^k(BE_8,\pi_{k-1}(F)).
\end{equation*}
By the previous
observation, this is isomorphic to $H^k(K(\BZ,4),\pi_{k-1}(F))$ in the relevant
range.

From the table above, there are three possible obstructions, living in
the groups
\begin{equation*}
H^3(M, \BZ_2), \: \:
H^9(M, \BZ), \: \:
H^{10}(M, \BZ_2).
\end{equation*}
The first of these we can eliminate immediately, since it is a pullback
from $H^3(BE_8,\BZ_2)$ but that group vanishes.
                                                                                
Next we check $H^9(K(\BZ,4), \BZ)=\BZ_3$ and 
$H^{10}(K(\BZ,4), \BZ_2)=\BZ_2+\BZ_2$.
These groups will yield two potential obstructions:
an element of $H^9(M, \BZ)$, pulled back from a class
in $H^9(K(\BZ, 4), \BZ)$,
and an element of $H^{10}(M, \BZ_2)$, pulled back from a class
in $H^{10}(K(\BZ,4), \BZ_2)$.
                                                                                
In principle, the universal obstruction in $H^9(K(\BZ,4), \BZ)$ can be
nonzero because it agrees
with the $k$-invariant of $KO$ at $p=3$.  Its name is ``Milnor's $Q_1$.''
It is a
cohomology operation $Q_1:H^n(-,\BZ) \rightarrow H^{n+5}(-,\BZ)$.
                                                                                
So, let us concentrate at $p=3$ for a moment.  The question is, 
does there exist a 10 dimensional manifold $M$ with a 4-dimensional
cohomology class $x$ on which $Q_1$ is non zero? 
It can be shown by a cobordism invariance argument
\cite{francispriv} that on any oriented 10-manifold
$M$, there is no such cohomology class.

Thus, so long as our 10-manifold $M$ is oriented,
the potential obstruction in $H^9(M, \BZ)$ always vanishes,
leaving us with only one potential obstruction to reductibility
of the structure group of the $E_8$ bundle,
living in $H^{10}(M, \BZ_2)$.
Unfortunately, this obstruction can sometimes be nonzero.
(Examples of oriented 10-manifolds with nonreducible $E_8$ bundles
are described in \cite{dmw}, albeit to different ends.)

Although we have been unable to find any prior references discussing
this obstruction, we have found some that came close to uncovering it.
For example, in \cite{edsymp}, Witten points out the necessity of
reducing $E_8$ to $\mathrm{Spin}(16)/\BZ_2$, and also looks
for obstructions, but only up to degree six:  he observes
that for compactifications to four dimensions, such a reduction is always
possible.  

\subsection{Target space interpretation}
So far we have discussed a technical issue that arises when 
trying to understand certain `exotic' $E_8$ bundles on a heterotic
string worldsheet.  Next, we shall discuss the interpretation of this
obstruction in the ten-dimensional supergravity.

For chiral fermions in dimension $8k+2$, it is known
\cite{edcmp}[p. 206] that the number of zero modes of the
chiral Dirac operator is a topological invariant mod 2.
(The number of zero modes of the nonchiral Dirac operator is a topological
invariant mod 4.)
In particular, since the ten-dimensional gaugino is a Majorana-Weyl spinor,
the number of posititive chirality gaugino zero modes is a 
topological invariant mod 2.  For $E_8$ bundles, this topological invariant
was discussed in \cite{dmw}[section 3], where it was labelled
$f(a)$ (where $a$ is the analogue of the Pontryagin invariant for
$E_8$ bundles).

Curiously, the element of $H^{10}(X, \BZ_2)$ that defines the
obstruction to reducing an $E_8$ bundle to a $\mathrm{Spin}(16)/\BZ_2$
bundle, is that same invariant \cite{hopkinspriv}.
In other words, the number of chiral gaugino zero modes
of the ten-dimensional Dirac operator is odd
precisely when the $E_8$ bundle cannot be reduced to
$\mathrm{Spin}(16)/\BZ_2$, and hence cannot be described
perturbatively on a heterotic string worldsheet.

This makes the current phenomenon sound analogous to the
anomaly in four-dimensional $SU(2)$ gauge theories with an odd number
of left-handed fermion doublets, described in \cite{edsu2}.
There, the anomaly could be traced to the statement that the
five-dimensional Dirac operator had an odd number of zero modes,
which translated into the statement that the relevant operator
determinant in the four-dimensional theory was not well-behaved
under families of gauge transformations.  There, however,
it was the Dirac operator in one higher dimension that had an odd
number of zero modes, whereas in the case being studied in this
paper it is the Dirac operator in ten dimensions, not eleven dimensions,
that has an odd number of zero modes.  Also, in the anomaly
studied in \cite{edsu2}, the fact that
$\pi_4(SU(2))$ is nonzero was crucial, 
whereas by contrast $\pi_{10}(E_8)$ vanishes.
In fact that last fact was used in \cite{edcmp}[p. 198] to argue that
there should not be any global gauge anomalies in heterotic
$E_8 \times E_8$ strings.

\section{Connections}\label{e8conn}
So far we have discussed reducibility of topological $E_8$ bundles
to $\mathrm{Spin}(16)/\BZ_2$ bundles, but to realize a given $E_8$
gauge field in standard heterotic string constructions, we must
also reduce the connection on the bundle, not just the bundle itself.

In particular, on a principal $G$-bundle, even a trivial principal
$G$-bundle, one can find connections with holonomy that fill out all
of $G$, and so cannot be understood as coming from connections on
any principal $H$-bundle for $H$ a subgroup of $G$.
It is easy to see this statement locally \cite{rthompriv}:
one can pick a connection
whose curvatures at points in a small open set generate the Lie algebra
of $G$, and then the local holonomy will generate (the identity component of)
$G$, and since our bundles are reducible (in fact, trivial) locally, one 
gets the desired result.

However, for our purposes it does not suffice to consider reducibility
of generic connections.
After all, for a perturbative vacuum of heterotic string theory, 
the connection must
satisfy some stronger conditions:  it must satisfy the Donaldson-Uhlenbeck-Yau
equation, the curvature must be of type $(1,1)$,
and it must satisfy anomaly cancellation.

However, even when the Donaldson-Uhlenbeck-Yau condition is satisfied,
it is still possible to have bundles with connection such that the
bundle is reducible but not the connection.
Examples of this were implicit in \cite{kcs},
which discussed how stability of bundles depends upon the metric.
Briefly, the K\"ahler cone breaks up into subcones, with a different
moduli space of bundles on each subcone.  Some stable irreducible bundles
will, on the subcone wall, become reducible.
This means that the holomorphic structure (and also the
holonomy of the connection) was generically irreducible,
but becomes reducible at one point.  For this to be possible at the
level of holomorphic structures means that the bundle was always
topologically reducible.  Thus, implicitly in \cite{kcs}
there were examples of topologically reducible bundles with
irreducible connections satisfying the Donaldson-Uhlenbeck-Yau condition.

We shall construct some examples on K3 surfaces of $E_8$ gauge fields
which satisfy all the conditions above for a perturbative heterotic
string vacuum, but which cannot be reduced to $\mathrm{Spin}(16)/\BZ_2$.

\subsection{Moduli spaces of flat connections}
As a quick warm-up, let us briefly study how the moduli space
of flat $E_8$ connections on $T^2$ arises in a heterotic compactification
on $T^2$.
The moduli space of flat $E_8$ connections on $T^2$ and one component
of the moduli space of flat $\mathrm{Spin}(16)/\BZ_2$ connections
both have the form $(T^2)^8/W$, where $W$ is the respective Weyl group.
However, $W(D_8) \subset W(E_8)$, and in fact $| W(E_8)/W(D_8)| = 135$,
so the component of the moduli space of flat $\mathrm{Spin}(16)/\BZ_2$
connections is a 135-fold cover of the moduli space of flat $E_8$ connections.

The projection to the moduli space of flat $E_8$ connections is induced
by T-dualities.  The discrete automorphism group (T-dualities) of the heterotic
moduli space includes a $O(\Gamma_8)$ factor, which acts as the
$E_8$ Weyl group action above.  When forming the moduli space,
we mod out by this factor, and so we get the moduli space of
flat $E_8$ connections, rather than that of $\mathrm{Spin}(16)/\BZ_2$
connections.

\subsection{Analysis of connections}
In this section we will construct
an example\footnote{We would like to thank R.~Thomas for an
extensive discussion of this matter in late March and April, 2006.} 
of an $E_8$ gauge field on a Calabi-Yau $X$
which cannot be reduced to $\mathrm{Spin}(16)/\BZ_2$, but which does
satisfy the conditions for a consistent perturbative vacuum,
namely
\begin{equation*}
F_{0,2} \: = \: F_{2,0} \: = \:
g^{i \overline{\jmath}} F_{i \overline{\jmath}} \: = \: 0
\end{equation*}
and that
\begin{equation*}
\mbox{Tr } F^2 \: - \: \mbox{Tr }R^2 
\end{equation*}
is cohomologous to zero.

To build this example, we use the fact that
$E_8$ contains a subgroup $\left( SU(5) \times SU(5) \right)/\BZ_5$.
This subgroup is not a subgroup of $\mathrm{Spin}(16)/\BZ_2$,
and so an $SU(5) \times SU(5) / \BZ_5$ gauge field whose holonomy
is all of the group is an example of an $E_8$ gauge field that cannot
be reduced to $\mathrm{Spin}(16)/\BZ_2$.
To construct such an $(SU(5) \times SU(5))/\BZ_5$ gauge field,
it suffices to construct an $SU(5) \times SU(5)$ gauge field,
then take the image under a $\BZ_5$ action (whose existence is
always guaranteed). 

The perturbative anomaly cancellation condition is stated simply as
a matching of $\mbox{Tr } F^2$ and $\mbox{Tr }R^2$ in cohomology,
but for general groups the precise interpretation of that statement
in terms of degree four characteristic classes.
For an $SU(5) \times SU(5)$ bundle, anomaly cancellation should be
interpreted as the statement
\begin{equation*}
c_2({\cal E}_1) \: + \: c_2({\cal E}_2) \: = \: c_2(TX)
\end{equation*}
where ${\cal E}_1$, ${\cal E}_2$ are principal $SU(5)$ bundles.

As a check of anomaly cancellation in this context, 
suppose that $SU(n)$ is a subgroup of $SU(5)$.
We can either embed the $SU(n)$ in $\mathrm{Spin}(16)/\BZ_2$,
and then build up a standard perturbative worldsheet, or
we can embed it in $SU(5)\times SU(5) / \BZ_5$, which does not
admit a perturbative description.
This gives two paths to $E_8$, but these two paths commute\footnote{We would
like to thank A.~Knutson for a helpful discussion of this matter at the
end of March 2006.  Also, note the automorphism exchanging the two $SU(5)$'s
does not extend to $E_8$, which can also be seen from the asymmetry of
the decomposition of the adjoint representation of $E_8$ under
the subgroup above.  Another way to see this is from
the fact that the $\BZ_5$ one quotients by is
not symmetric under such a switch.}.

A careful reader might point out another subtlety in the
statement of anomaly cancellation.
For example, the degree four characteristic class of an $SU(n)/\BZ_n$
bundle obtained from an $SU(n)$ bundle ${\cal E}$ can be
naturally taken to be\footnote{Alternatively,we can get the same result from the fact thatthe trace in the adjoint rep of $SU(n)$
is $2n$ times the trace in the fundamental rep \cite{erler},
which is also twice the dual Coxeter number.
}
$c_2(\mbox{End}_0 {\cal E}) = 2n c_2({\cal E})$,
so in the case above there could plausibly be extra numerical factors.
In any event, our methods are sufficiently robust that
such modifications of the anomaly cancellation condition will not
change the fact that there exist families of examples\footnote{If the reader
objects that a wandering factor of $5$ or $10$, as might be expected
in some interpretations of $SU(5)^2/\BZ_5$, would make examples
on K3's difficult, the quintic threefold has $c_2$ divisible by
5, in fact $c_2 = 10 H^2$, and there exist further examples there.}.
Put another way, nonreducible
connections are common, not rare or unusual.

We need to find a bundle with connection that not only satisfies 
anomaly cancellation, but also the Donaldson-Uhlenbeck-Yau condition.
By working with $SU(n)$ gauge fields, we can translate such questions about
connections
into algebraic geometry questions.  
In particular, the requirement that
the gauge field satisfy the Donaldson-Uhlenbeck-Yau equation becomes
the requirement that the corresponding holomorphic rank 5 vector bundle
be stable.

Ordinarily, checking stability can be rather cumbersome, but there is
an easy way to build examples sufficient for our purposes. 
We can build holomorphic vector bundles on elliptic fibrations with
section using the techniques of \cite{fmw,bjps}.
(See also {\it e.g.} \cite{bjorn1,bjorn2,bjorn3,bjorn4,ovrut1,ovrut2} for some
more modern applications of the same technology.)
Furthermore, these bundles are automatically stable (for metrics in the
right part of the K\"ahler cone).
One must specify a (spectral) cover of the base of the fibration, plus a line
bundle on that cover.

Following the conventions of \cite{bjps},
to describe an $SU(r)$ bundle on an elliptic K3 with section we use
a spectral cover describing an $r$-fold cover of the base of the fibration.
The spectral cover will be in the class $| r \sigma + k f|$
where $\sigma$ is the class of the section and $f$ is the class
of the fiber, and $k$ is the second Chern class of the bundle
\cite{bjps}[p. 5].

Furthermore, there is a line bundle that must be specified on that cover,
and it can be shown \cite{bjps} that that line bundle must
have degree $-(r+g-1)$, where $g = rk - r^2 + 1$ is the genus of
the spectral cover (as it is a cover of ${\bf P}^1$, it is some
Riemann surface).  If the spectral curve is reduced and irreducible
then the corresponding bundle will be stable; Bertini's theorem
implies that such curves exist in the linear system.

In the present case, we want a holomorphic vector bundle of
rank $5$, $c_1=0$, $c_2=12$.  The spectral cover that will produce
such a result is in the linear system
$|5 \sigma + 12 f|$.  The genus of such a curve is $36$, and the 
line bundle has degree $-40$.
The dimension of the  moduli space of spectral data is then $2 \cdot 36
= 72$.

So far we have established the existence of stable $SU(5) \times
SU(5)$ bundles satisfying all the conditions for a consistent
perturbative vacuum; we still need to demonstrate that the holonomy
of the connection cannot be reduced below $SU(5) \times SU(5)$.
To do this we can apply the recent work \cite{kollar}, which says
that it is sufficient for each factor to be irreducible and to have
irreducible second symmetric power.  As this will be generically
true \cite{donagipriv}, we see that the holonomy cannot be reduced
below $SU(5) \times SU(5)$, and so by projecting along a $\BZ_5$
automorphism we have a family of $(SU(5) \times SU(5))/\BZ_5$ 
bundles with the desired properties.

Thus, using the embedding of $(SU(5) \times SU(5))/\BZ_5$ in
$E_8$, we now have a family of $E_8$ bundles with connection on 
K3's which satisfy all the requirements for a consistent perturbative
vacuum, but which cannot be reduced to $\mathrm{Spin}(16)/\BZ_2$,
and so cannot be described with standard constructions of heterotic
strings.

\subsection{Low energy theory}
Compactification on a bundle with structure group $(SU(5) \times
SU(5)) / \BZ_5$ breaks the $E_8$ to a mere $\BZ_5$ -- 
the commutant in $E_8$ is $\BZ_5$.
Similarly, if one were to compactify on a bundle with structure
group $\mathrm{Spin}(16)/\BZ_2$, the commutant inside $E_8$
is $\BZ_2$.

If it were the case that the low-energy theory in any $E_8$ bundle
not describable on the worldsheet had gauge group only a finite
group, then this might not be considered very interesting.
However, there are other examples of subgroups of $E_8$ whose
commutant has rank at least one, and which cannot be
embedded in $\mathrm{Spin}(16)/\BZ_2$.

For example, the group $(E_7 \times U(1))/\BZ_2$
is a subgroup of $E_8$ (that sits inside the $(E_7 \times SU(2))/\BZ_2$
subgroup of $E_8$) which has commutant $U(1)$,
and is not a subgroup of $\mathrm{Spin}(16)/\BZ_2$.

For another example, $(E_6 \times SU(3))/\BZ_3$ is a subgroup
of $E_8$, and so its $E_6$ subgroup has commutant $SU(3)$,
but $E_6$ cannot be embedded in $\mathrm{Spin}(16)/\BZ_2$.
To see this, note that if $E_6$ could be embedded in $\mathrm{Spin}(16)/\BZ_2$,
then the Lie algebra $so(16)$ would have an $e_6$ subalgebra,
and since there is a 16-dimensional representation of $so(16)$,
that means $e_6$ would have a possibly reducible nontrivial
16-dimensional representation
as well, just from taking the subalgebra described by some of the
$16 \times 16$ matrices describing $so(16)$.  However, the smallest
nontrivial representation of $e_6$ is 27-dimensional, a contradiction.
(Note this is closely related to but distinct from the standard
embedding for Calabi-Yau three-folds:
the $SU(3)$ subgroup of $( E_6 \times SU(3) )/\BZ_3$ {\it does}
sit inside $\mathrm{Spin}(16)/\BZ_2$, unlike the $E_6$.)

\section{F theory duals and the existence of perturbative realizations}\label{fthydual}
So far we have argued that there exist some bundles with connection that
cannot be realized using the standard description of heterotic $E_8 \times
E_8$ strings.  Does that mean that they do not arise in string theory?
Such questions are important to the landscape program, for example,
where one of the current issues involves understanding which
backgrounds admit UV completions \cite{bankstalk,vafaswamp}.

Some insight into this question can be made with F theory duals.
For example, \cite{paulrec}[section 2.3] describes an F theory dual
to a heterotic compactification in which the bundle with connection
has structure group $(E_7 \times U(1))/\BZ_2$, and so cannot
be realized with the standard construction of heterotic strings.

Such examples tell us that at least some of these bundles with connection
can nevertheless be realized within string theory.

More abstract considerations lead one to the same conclusion.
Imagine starting with a bundle with connection reducible to
$\mathrm{Spin}(16)/\BZ_2$, and deforming to an $E_8$ bundle with
connection that is not reducible.  Since the adjoint representation of
$E_8$ decomposes into the adjoint and a chiral spinor representation of
$\mathrm{Spin}(16)/\BZ_2$, the deformation described would involve
giving a vacuum expectation value to a spinor.
This sounds reminiscent of describing Ramond-Ramond fields in type II
strings with nonzero
vacuum expectation values.  In the case of type II strings,
giving those fields vacuum expectation values involved formally adding
terms to the lagrangian coupled to the superconformal ghosts,
which is problematic, and is the reason that Ramond-Ramond field vevs
are problematic in basic formulations of type II strings.
In a heterotic string, however, giving a vev to a gauge spinor does
{\it not} involve coupling to superconformal ghosts, unlike the type II
case, so there is no obstruction in principle.
Thus, from this consideration, one is led to believe that
$E_8$ bundles with connection that cannot be reduced to 
$\mathrm{Spin}(16)/\BZ_2$ should nevertheless define 
well-behaved CFT's, even though they cannot
be described within traditional heterotic worldsheet constructions.

In the remainder of this paper we will describe alternative constructions
of perturbative heterotic strings which can explicitly
realize more general $E_8$ bundles with
connection.  First, in the next section we will describe how 
subgroups other than $\mathrm{Spin}(16)/\BZ_2$ can be used to build
$E_8$ in ten dimensions, and will check by comparing modular forms
that corresponding current algebra constructions realize all of the
degrees of freedom of the left-moving part of the standard constructions.
To make such constructions practical in less than ten dimensions,
however, one needs suitable technology for fibering current algebras
over a base, and so we introduce ``fibered WZW models,'' which will
enable us to fiber a current algebra for any group at any level
over a base, using a principal bundle with connection to define the
fibering.

\section{Alternative constructions of 10d heterotic strings}\label{alt10d}
The reader might ask whether the heterotic string could be formulated
in some alternative fashion that might be more amenable to some of the
constructions above.  For example, might it be possible to formulate
a worldsheet string with, for each $E_8$, two sets of five complex fermions,
realizing the $E_8$ from $(SU(5) \times SU(5) ) / \BZ_5$?
Unfortunately, two sets of five complex fermions would have a 
$U(5) \times U(5)$ global symmetry, and if we try to gauge
each $U(1)$ on the worldsheet, we would encounter a $U(1)^2$ anomaly
which would force $c_2$ of each bundle to vanish separately.

Instead, we are going to take an alternative approach to this issue.
We are going to develop a notion of fibered current algebras,
realized by fibered WZW models,
which will allow us to realize current algebras at any level
and associated to any group $G$, fibered nontrivially over
any compactification manifold. 
The standard $E_8 \times E_8$ heterotic
string construction is, after all, one realization of a fibered
$E_8 \times E_8$ current algebra at level 1; our technology will
enable us to talk about fibering $G$-current algebras at level $k$.

Before doing that, however,
we will check to what extent subgroups of $E_8$ other than
$\mathrm{Spin}(16)/\BZ_2$ can be used to build up the left-moving
$E_8$ partition function in ten dimensions.
For example, one could take a pair of $SU(5)$ current algebras,
then perform a $\BZ_5$ orbifold (replacing the ``left-moving GSO''
used to build $\mathrm{Spin}(16)/\BZ_2$ from a $\mathrm{Spin}(16)$
current algebra in the usual construction)
so as to get
an $( SU(5) \times SU(5) )/\BZ_5$ global symmetry on the
worldsheet, or take an $SU(9)$ global symmetry and perform a
$\BZ_3$ orbifold to get an $SU(9)/\BZ_3$ global symmetry.
Both $(SU(5)\times SU(5))/\BZ_5$ and $SU(9)/\BZ_3$
are subgroups of $E_8$,
and we will find that such alternative subgroups correctly reproduce
the $E_8$ partition function, and so give alternative constructions
of the $E_8$ current algebra in ten dimensions.
At the level of characters and abstract affine algebras, the idea
that $E_8$ can be built from other subgroups has appeared previously
in \cite{kacsan}; we shall review some pertinent results and also
describe how those character decompositions are realized physically
in partition functions, via orbifold twisted sectors.

First, let us recall how $E_8$ is built from $\mathrm{Spin}(16)/\BZ_2$
in ten dimensions.
The adjoint representation of $E_8$ decomposes as
\begin{equation}    \label{e8spin16}
{\bf 248} \: = \: {\bf 120} \: + \: {\bf 128}
\end{equation}
under $\mathrm{Spin}(16)/\BZ_2$.
At the level of ordinary Lie algebras, we get the elements of the $E_8$
Lie algebra from the adjoint plus a spinor representation of 
$\mathrm{Spin}(16)/\BZ_2$, and assigning them suitable commutation relations.
At the level of WZW conformal families, we could write
\begin{equation*}
[{\bf 1}] \: = \: [{\bf 1}] \: + \: [{\bf 128}]
\end{equation*}
which implicitly includes equation~(\ref{e8spin16}) as a special case,
since the (adjoint-valued) currents are non-primary descendants of the
identity operator.  
That statement about conformal families implies a statement about
characters of the corresponding affine Lie algebras, namely that
\begin{equation}   \label{spin16base}
\chi_{E_8}({\bf 1},q) \: = \: \chi_{Spin(16)}({\bf 1}, q)
\: + \: \chi_{Spin(16)}({\bf 128}, q)
\end{equation}
where \cite{gswv1}[section 6.4.8]
\begin{equation*}
\chi_{E_8}({\bf 1}, q) \: = \: \frac{E_2(q)}{\eta(q)^8}
\end{equation*}
and where $E_2(q)$ is the degree four Eisenstein modular form
\begin{equation*}
\begin{split}
E_2(q) = \: & 1 \: + \: 240 \sum_{m=1}^{\infty} \sigma_3(m) q^m \\
= \: & 1 \: + \: 240 \left[ q \: + \: (1^3 + 2^3) q^2 \: + \: (1^3 + 3^3) q^3
\: + \: \cdots \right] \\
= \: & 1 \: + \: 240 q \: + \: 2160 q^2 \: + \: 6720 q^3 \: + \:
17520 q^4 \: + \: 30240 q^5 \: + \: 60480 q^6 \: + \: \cdots
\end{split}
\end{equation*}
with
\begin{equation*}
\sigma_3(m) \: = \: \sum_{d | m} d^3
\end{equation*}
The identity~(\ref{spin16base}) is discussed in for example
\cite{gswv1}[section 6.4] and \cite{gannonlam1}[eqn (3.4a)].  
The $\BZ_2$ orbifold plays a crucial role in the expression above.
Without the $\BZ_2$ orbifold, we would only consider the single
conformal family $[{\bf 1}]$ and the single character
$\chi_{Spin(16)}({\bf 1}, q)$.   The $[{\bf 128}]$ 
arises from the $\BZ_2$ orbifold twisted sector.
(The fact that the twisted sector states are still representatives of
the same affine Lie algebra as the untwisted sector states, despite
being in a twisted sector, is a consequence of the fact that the 
orbifold group action preserves the currents -- it acts on the center
of the group, preserving the algebra structure.)

Next, we shall check to what extent other subgroups of $E_8$ can be
used to duplicate the same left-moving degrees of freedom.

\subsection{ Some maximal-rank subgroups}
In this subsection, we shall argue that the left-moving
$E_8$ degrees of freedom can be reproduced by using the
maximal-rank
$SU(5)^2/\BZ_5$ and $SU(9)/\BZ_3$ subgroups of $E_8$,
in place of $\mathrm{Spin}(16)/\BZ_2$.
Just as for $\mathrm{Spin}(16)/\BZ_2$, the finite group quotients
will be realized by orbifolds and will play a crucial role.
At the level of characters of affine algebras, the ideas have
appeared previously in {\it e.g.} \cite{kacsan}, but we shall also
explain how those character decompositions are realized physically
in partition functions.
For more information on determining such finite group quotients,
see appendix~\ref{gpthy}.

First, let us check central charges.
From \cite{diFranc}[section 15.2],
the central charge of a bosonic WZW model at level $k$ is
\begin{equation*}
\frac{ k \, \mbox{dim }G }{ k + C }
\end{equation*}
where $C$ is the dual Coxeter number.
For the case of $G = SU(N)$, $\mbox{dim }G = N^2 - 1$
and $C=N$ (see {\it e.g.} \cite{ps}[p. 502]), hence the central charge
of the bosonic $SU(N)$ WZW is
\begin{equation*}
\frac{ k(N^2 - 1) }{k + N}
\end{equation*}
For $k=1$, this reduces to $N-1$.  Thus, the $SU(5)$ current algebra
at level 1 has central charge 4, and the $SU(9)$ current algebra has
central charge 8.
In particular, this means that the $SU(5)\times SU(5)$ current
algebra at level 1
has central charge $4+4=8$,
just right to be used in critical
heterotic strings to build an $E_8$.  Similarly,
the $SU(9)$ current algebra at level 1 has central charge $8$, 
also just right to be used in critical heterotic strings to build an $E_8$.

Similarly, for $E_6$, $E_7$, $E_8$, the dual Coxeter numbers are 12, 18, 30,
respectively, and it is easy to check that at level 1, each
current algebra has central charge equal to 6, 7, 8, respectively.
More generally, for ADE groups, the level 1 current algebras have central
charge equal to the rank of the group.

For $SU(5)$, the integrable representations (defining WZW primaries)
are ${\bf 5}$, ${\bf 10} = \Lambda^2 {\bf 5}$,
${\bf \overline{10}} = \Lambda^3 {\bf 5}$, and
${\bf \overline{5}} = \Lambda^4 {\bf 5}$.
The fusion rules obeyed by the WZW conformal families have
the form
\begin{equation*}
\begin{split}
[{\bf 5}] \times [{\bf 5}] = \: & [{\bf 10}] \\
{[{\bf 5}]} \times [{\bf \overline{5}}] = \: & [{\bf 1}] \\
{[{\bf \overline{10}}]} \times [{\bf \overline{5}}] = \: & [{\bf 10}] \\
{[{\bf 10}]} \times [{\bf \overline{5}}] = \: & [{\bf 5}] \\
{[{\bf \overline{10}}]} \times [{\bf \overline{10}}] = \: & [{\bf 5}] \\
{[{\bf \overline{10}}]} \times [{\bf 10}] = \: & [{\bf 1}] 
\end{split}
\end{equation*}
The adjoint representation of $E_8$ decomposes under $SU(5)^2/\BZ_5$ as
\cite{slansky}
\begin{equation*}
{\bf 248} \: = \: ({\bf 1}, {\bf 24}) \: + \: ({\bf 24},{\bf 1}) \: + \:
({\bf 5},{\bf \overline{10}}) \: + \: ({\bf \overline{5}},{\bf 10})
\: + \: ({\bf 10},{\bf 5}) \: + \: ({\bf \overline{10}},{\bf \overline{5}})
\end{equation*}
from which one would surmise that the corresponding statement about
conformal families is
\begin{equation}   \label{su5conffam}
[{\bf 1}] \: = \: [{\bf 1},{\bf 1}] \: + \:
[{\bf 5},{\bf \overline{10}}] \: + \:
[{\bf \overline{5}},{\bf 10}] \: + \:
[{\bf 10},{\bf 5}] \: + \:
[{\bf \overline{10}},{\bf \overline{5}}]
\end{equation}
which can be checked by noting that the right-hand side above squares
into itself under the fusion rules.

Next, we shall check partition functions, which will provide the
conclusive demonstration that
the $E_8$ of a ten-dimensional heterotic string
can be built from $( SU(5) \times SU(5) )/\BZ_5$ 
instead of $\mathrm{Spin}(16)/\BZ_2$.

The character of the identity representation of $SU(5)$ is
\begin{equation*}
  \chi_{SU(5)}({\bf 1}, q) = \frac{1}{\eta(\tau)^{4}}\sum_{\vec{m}\in\BZ^{4}}
       q^{(\sum m_i^2 +(\sum m_i)^2)/2}
\end{equation*}
Taking modular transformations, the characters of the other
needed integrable representations are
\begin{equation*}
\chi_{SU(5)}({\bf 5}, q) = \frac{1}{\eta(\tau)^4} \sum_{{\vec{m}\in\BZ^4, \:  
\sum m_i=1 \bmod
5}}
q^{(\sum m_i^2 -\frac{1}{5}(\sum m_i)^2)/2}
\end{equation*}
and
\begin{equation*}
\chi_{SU(5)}({\bf 10}, q) = \frac{1}{\eta(\tau)^4} \sum_{{\vec{m}\in\BZ^4, \: 
\sum m_i=2 \bmod 5}}
q^{(\sum m_i^2 -\frac{1}{5}(\sum m_i)^2)/2}
\end{equation*}
The remaining two characters (given by $\sum m_i = 3,4 \bmod 5$) are equal
 to these, by taking $\vec{m}\to -\vec{m}$.

Now, we need to verify that
\begin{equation}   \label{e8su5chars}
  \chi_{E_8}({\bf 1}, q) = \chi_{SU(5)}({\bf 1}, q)^2 + 4 \chi_{SU(5)}(
{\bf 5}, q)\, \chi_{SU(5)}({\bf 10}, q)
\end{equation}
which corresponds to equation~(\ref{su5conffam}) for the conformal families.
This character decomposition, along with character decompositions for
other subgroups, has appeared previously in \cite{kacsan}, but since it
plays a crucial role in our arguments, we shall explain in detail why it is
true, and then explain how it is realized physically in partition functions.
The $E_8$ character is given by \cite{gswv1}[section 6.4.8]
\begin{equation*}
\chi_{E_8}({\bf 1},q) \: = \: \frac{E_2(q)}{\eta(\tau)^8}
\end{equation*}
where $E_2(q)$ denotes the relevant Eisenstein series.
The $\BZ_5$ orbifold is implicit here -- $\chi({\bf 1},q)^2$ arises
from the untwisted sector, and each of the four
$\chi({\bf 5},q)\chi({\bf 10},q)$'s arises from a twisted sector.
(As for $\mathrm{Spin}(16)/\BZ_2$, since the orbifold action preserves
the currents, the twisted sector states must form a well-defined module
over the (unorbifolded) affine Lie algebra.)

Ample numerical evidence for equation~(\ref{e8su5chars})
is straightforward to generate.
For example:
\begin{equation*}
\begin{split}
\eta(\tau)^4 \chi_{SU(5)}({\bf 1}, q) = \: & 1 \: + \: 20 q \: + \: 30 q^2 \: + \:
60 q^3 \: + \: 60 q^4 \: + \: 120 q^5 \: + \: 40 q^6 \: + \:
180 q^7 \\
& \: + \: 150 q^8 \: + \: 140 q^9 \: + \: 130 q^{10} \: + \:
240 q^{11}  \: + \: 180 q^{12} \: + \: 360 q^{13} \: + \: \cdots \\
\eta(\tau)^4 \chi_{SU(5)}({\bf 5}, q) = \: & 5 q^{2/5} \: + \: 30 q^{7/5} \: + \:
30 q^{12/5} \: + \: 80 q^{17/5} \: + \: 60 q^{22/5} \: + \:
100 q^{27/5} \\
& \: + \: 104 q^{32/5} \: + \: 168 q^{37/5} \: + \:
54 q^{42/5} \: + \: 206 q^{47/5} \: + \: 168 q^{52/5} \\
& \: + \: 172 q^{57/5} \: + \: 140 q^{62/5} \: + \:
270 q^{67/5} \: + \:
153 q^{72/5} \: + \: \cdots \\
\eta(\tau)^4 \chi_{SU(5)}({\bf 10}, q) = \: & 10 q^{3/5} \: + \: 25 q^{8/5} \: + \:
60 q^{13/5} \: + \: 35 q^{18/5} \: + \: 110 q^{23/5} \: + \:
90 q^{28/5}  \\
& \: + \: 120 q^{33/5} \: + \: 96 q^{38/5} \: + \:
198 q^{43/5} \: + \: 98 q^{48/5} \: + \: 244 q^{53/5} \\
& \: + \: 126 q^{58/5} \: + \: 192 q^{63/5} \: + \:
208 q^{68/5} \: + \:
300 q^{73/5} \: + \: \cdots
\end{split}
\end{equation*}
Putting this together, we find
\begin{multline*}
\lefteqn{ \eta(\tau)^8\left( \chi_{SU(5)}({\bf 1}, q)^2 \: + \: 
4 \chi_{SU(5)}({\bf 5}, q) \, \chi_{SU(5)}({\bf 10}, q)
\right) \: = \: } \\
1 \: + \: 240 q \: + \: 2160 q^2 \: + \: 6720 q^3 \: + \:
17520 q^4 \: + \: 30240 q^5 \: + \: 60480 q^6 \: + \: \cdots
\end{multline*}
which are precisely the first few terms of the appropriate Eisenstein series
$E_2(q)$, numerically verifying the prediction~(\ref{e8su5chars}).

More abstractly, the equivalence can be proven as follows\footnote{This
argument is due to E.~Scheidegger, and we would like to thank him
for allowing us to print it here.}.
In the notation of \cite{gannonlam1}, we need to relate the theta
function of the $E_8$ lattice to a product of theta functions for
$SU(5)$ lattices.  Briefly, first one argues that
\begin{equation*}
\Theta(E_8) \: = \: \Theta( \{ A_4, A_4 \}[1,2])
\end{equation*}
Using \cite{gannonlam1}[eqns (1.1), (1.5)], this can be written as
\begin{equation*}
\Theta\left( \bigcup_{i=1}^5 [ig]\{A_4, A_4 \} \right) \: = \:
\sum_{i=1}^5 \Theta\left( [ig]\{ A_4, A_4 \} \right)
\end{equation*}
where $g$ denotes the generator of the $\BZ_5$ action
(shift by 1 on first $A_4$, shift by 2 on second).
Using \cite{gannonlam1}[eqn (1.4)], this can be written as
\begin{equation*}
\sum_{i=1}^5 \Theta([ig]A_4) \Theta([ig]A_4)
\: = \: \sum_{i=1}^5 \Theta([i]A_4) \Theta([2i]A_4)
\end{equation*}
Using the symmetry
\begin{equation*}
\Theta([5-i]A_4) \: = \: \Theta([i]A_4)
\end{equation*}
the result then follows after making the identifications
\begin{equation*}
\eta(\tau)^4 \chi({\bf 1},q) \: = \: \Theta(A_4), \: \: \:
\eta(\tau)^4 \chi({\bf 5},q) \: = \: \Theta([1]A_4), \: \: \:
\eta(\tau)^4 \chi({\bf 10},q) \: = \: \Theta([2]A_4)
\end{equation*}

Merely verifying the existence of a character decomposition does not suffice
to explain how this can be used in alternative constructions of heterotic
strings -- one must also explain how that character decomposition is
realized physically.  In the case of $\mathrm{Spin}(16)/\BZ_2$, the
two components of the character decomposition were realized physically
as the untwisted and twisted sectors of a $\BZ_2$ orbifold
of a $\mathrm{Spin}(16)$ current algebra.  That orbifold structure
precisely correlates with the group-theoretic fact that the
subgroup of $E_8$ is $\mathrm{Spin}(16)/\BZ_2$ and not
$\mathrm{Spin}(16)$ or $SO(16)$ -- the finite group factor that one gets
from the group theory of $E_8$, appears physically as the orbifold of the
current algebra that one needs in order to reproduce the correct
character decomposition.

There is a closely analogous story here.
Group-theoretically, the subgroup of $E_8$ is not $SU(5) \times SU(5)$
but rather $( SU(5) \times SU(5) )/\BZ_5$, and so one should expect
that a $\BZ_5$ orbifold of the $SU(5)\times SU(5)$ current algebra
should appear.  Indeed, that is precisely what happens.
If we only considered an $SU(5)\times SU(5)$ current algebra without 
an orbifold, the only contribution to the heterotic partition function
would be from the characters $\chi_{SU(5)}({\bf 1},q)^2$,
which would not reproduce the $E_8$ character.
In order to realize the complete $E_8$ character decomposition, we need
more, and the extra components of the character decomposition are realized
in twisted sectors of a $\BZ_5$ orbifold, the same $\BZ_5$ arising
in group-theoretic considerations.
Each $\chi_{SU(5)}({\bf 5}, q) \chi_{SU(5)}({\bf 10}, q)$ arises
in a twisted sector.  The individual ${\bf 5}$, ${\bf 10}$,
${\bf \overline{5}}$, and ${\bf \overline{10}}$ are not invariant under
the $\BZ_5$, but the products $({\bf 5}, {\bf \overline{10}})$,
$({\bf \overline{5}}, {\bf 10})$, $({\bf 10}, {\bf 5})$,
$({\bf \overline{10}}, {\bf \overline{5}})$ {\it are} invariant
under the $\BZ_5$ orbifold, as discussed in appendix~\ref{gpthy}.

For $SU(9)/\BZ_3$, there is an analogous\footnote{At the level of
character decompositions, this and other examples are discussed in {\it e.g.}
\cite{kacsan}.} story.
The adjoint representation of $E_8$ decomposes as \cite{slansky}
\begin{equation*}
{\bf 248} \: = \: {\bf 80} \: + \: {\bf 84} \: + \:
{\bf \overline{84}}
\end{equation*}
and so proceeding as before the conformal families of $E_8$, $SU(9)$ should
be related by
\begin{equation}    \label{su9conffam}
[ {\bf 1} ] \: = \: [ {\bf 1} ] \: + \: [ {\bf 84} ] \: + \:
[ {\bf \overline{84}} ]
\end{equation}
(which includes the decomposition above as a special case as the
currents in the current algebra are descendants of the identity).
The relevant $SU(9)$, level 1, characters are given by
\begin{equation*}
   \chi_{SU(9)}({\bf 1}, q) = \frac{1}{\eta(\tau)^8}
     \sum_{\vec{m}\in \BZ^8} q^{(\sum m_i^2 +(\sum m_i)^2)/2}
\end{equation*}
and
\begin{equation*}
   \chi_{SU(9)}({\bf 84}, q) = \frac{1}{\eta(\tau)^8}
     \sum_{{\vec{m}\in \BZ^8, \:
           \sum m_i =3 \bmod 9}} q^{(\sum m_i^2 -\frac{1}{9}(\sum m_i)^2)/2}
\end{equation*}
(The character for ${\bf \overline{84}}$ is identical.)
Then, from equation~(\ref{su9conffam}) it should be true that
\begin{equation*}
   \chi_{E_8}({\bf 1}, q) = \chi_{SU(9)}({\bf 1}, q) \: + \: 
2 \chi_{SU(9)}({\bf 84}, q)
\end{equation*}
This identity is proven in \cite{gannonlam1}[table 1].
The same statement is also made for lattices in
\cite{lerchelattice}[section A.3, p. 109] and
\cite{go}[eqn (8.12)], and of course also appeared in 
\cite{kacsan}.

Again, it is important to check that this character decomposition
really is realized physically in a partition function,
and the story here closely mirrors the $(SU(5)\times SU(5))/\BZ_5$
and $\mathrm{Spin}(16)/\BZ_2$ cases discussed previously.
Group-theoretically, the subgroup of $E_8$ is $SU(9)/\BZ_3$
and not $SU(9)$ or $SU(9)/\BZ_9$, so one would expect that we need
to take a $\BZ_3$ orbifold of the $SU(9)$ current algebra.
Indeed, if we did not take any orbifold at all, and only coupled
the $SU(9)$ current algebra by itself, then the only contribution
to the heterotic partition function would be from the
character $\chi_{SU(9)}({\bf 1},q)$, which does not suffice to reproduce
the $E_8$ character.  Instead, we take a $\BZ_3$ orbifold,
and each of the two characters $\chi_{SU(9)}({\bf 84}, q)$,
$\chi_{SU(9)}({\bf \overline{84}},q)$ appears in a $\BZ_3$ orbifold
twisted sector.  Taking those orbifold twisted sectors into account
correctly reproduces the $E_8$ character decomposition within the
heterotic partition function.

\subsection{A non-maximal-rank subgroup }
So far we have discussed how $E_8$ can be built from maximal-rank subgroups.

Somewhat surprisingly, on the level of characters, it appears that 
one can build it from non-maximal-rank subgroups also. 
We will discuss the case of $G_2 \times F_4$.
Although it satisfies many highly nontrivial checks, unfortunately
we will eventually conclude that it cannot be used, unlike the
maximal-rank subgroups discussed so far.

First, we should mention that the construction of the ordinary
Lie group $E_8$ from $G_2 \times F_4$ is described in \cite{adams}[chapter 8].
Very roughly, the idea is that if one takes $\mathrm{Spin}(16)$
and splits it into $\mathrm{Spin}(7) \times \mathrm{Spin}(9)$,
then $G_2 \subset \mathrm{Spin}(7)$ and $F_4 \subset \mathrm{Spin}(9)$.
Under the $g_2 \times f_4$ subalgebra, the adjoint representation of $e_8$
decomposes as \cite{slansky}
\begin{equation}   \label{e8g2f4}
{\bf 248} \: = \: ({\bf 14}, {\bf 1}) \: + \: ({\bf 1}, {\bf 52})
\: + \: ({\bf 7}, {\bf 26})
\end{equation}

The commutant of $G_2 \times F_4$ in $E_8$ has rank zero.
One way to see this is from the construction outlined above, but a simpler
way is from the decomposition of the adjoint representation of $E_8$:
if the commutant had rank greater than zero, then the adjoint of the
commutant would secretly appear in the decomposition of the adjoint of
$E_8$, as a set of singlets, but there are no singlets in the $E_8$
adjoint decomposition, and so the commutant must have rank zero.

Thus, even though $G_2 \times F_4$ is not of maximal rank, its commutant
in $E_8$ can be no more than a finite group.
This may sound a little surprising to some readers, but is in fact
a relatively common occurrence in representation theory.
For example, a dimension $n$ representation of $SU(2)$ embeds
$SU(2)$ in $SU(n)$, and has rank zero commutant inside $SU(n)$,
even though $SU(2)$ is not a maximal-rank subgroup.
This is a consequence of Schur's lemma.

We are going to discuss whether the $E_8$ degrees of freedom can be
described by this non-maximal-rank subgroup, namely $G_2 \times F_4$.
As one initial piece of evidence,
the fact stated above that the commutant of $G_2 \times F_4$ in $E_8$
has rank zero is consistent.  After all,
if it is possible to describe all of the $E_8$ current algebra using 
$G_2\times F_4$ on the internal space, then there will be no
left-moving worldsheet degrees of freedom left over to describe
any gauge symmetry in the low-energy compactified heterotic theory.
That can only be consistent if the commutant has rank zero,
{\it i.e.}, if there is no low-energy gauge symmetry left over to
describe.

Next, let us check that the central charges of the algebras work out
correctly.  The dual Coxeter number of $G_2$ is $4$ and that of $F_4$
is $9$, so the central charge of the $G_2$ algebra at level 1 is
$14/5$ and that of the $F_4$ algebra at level 1 is $52/10$,
which sum to $8$, the same as the central charge of the $E_8$ algebra
at level 1.

Both $G_2$ and $F_4$ affine algebras at level one have only two\footnote{
This is a short exercise using \cite{slansky}, let us briefly outline
the details for $G_2$.  The condition for a representation with highest
weight $\lambda$ to be integrable at level $k$ is $2 \psi \cdot \lambda / \psi^2
\leq k$, where $\psi$ is the highest weight of the adjoint representation.
Using \cite{slansky} tables 7 and 8, a representation of $G_2$ with
Dynkin labels $(a,b)$ has $2 \psi \cdot \lambda / \psi^2 = 2a+b$,
where $a$, $b$ are nonnegative integers, and so can only be $\leq 1$
when $a=0$ and $b$ is either $0$ or $1$, which gives the
${\bf 1}$ and ${\bf 7}$ (\cite{slansky}[table 13]) representations
respectively.
}
integrable representations:
\begin{equation*}
\begin{array}{cc}
G_2: & [{\bf 1}], [{\bf 7}] \\
F_4: & [{\bf 1}], [{\bf 26}]
\end{array}
\end{equation*}
The conformal weights of the primary fields are, respectively, $h_{7}= \tfrac{2}{5}$ and $h_{26}= \tfrac{3}{5}$
So, our proposed  decomposition of $E_8$ level 1 (which has only one integrable
representation) 
\begin{equation*}
[{\bf 1}] \: = \: [{\bf 1},{\bf 1}] \: + \: [{\bf 7},{\bf 26}]
\end{equation*}
does, indeed, reproduce the correct central charge and the conformal weights 
and multiplicity of currents.

Under modular transformations,
\begin{equation}\label{g2f4chardecomp}
   \chi_{E_{8}}({\bf 1},q) = \chi_{G_{2}}({\bf 1},q)\chi_{F_{4}}({\bf 1},q)+ \chi_{G_{2}}({\bf 7},q)\chi_{F_{4}}({\bf 26},q)
\end{equation}
transform identically. To see this, note that the fusion rules of $G_{2}$ 
and $F_{4}$ at level 1 are, respectively,
\begin{equation}\label{g2f4fusion}
\begin{array}{cc}
G_2: & [{\bf 7}] \times [{\bf 7}] \: = \: [{\bf 1}] + [{\bf 7}]\\
F_4: & [{\bf 26}] \times [{\bf 26}] \: = \: [{\bf 1}] + [{\bf 26}]
\end{array}
\end{equation}
The modular S-matrix (for both $G_{2}$ and $F_{4}$) is
\begin{equation}
  S = \frac{1}{\sqrt{2}}\begin{pmatrix}
  \sqrt{1-1/\sqrt{5}} &   \sqrt{1+1/\sqrt{5}} \\
  \sqrt{1+1/\sqrt{5}} & - \sqrt{1-1/\sqrt{5}}
  \end{pmatrix}
\end{equation}
which, in both cases, satisfies $S^{2}= (ST)^{3}=1$ and 
$N_{ijk}= \sum_{m}\tfrac{S_{im}S_{jm}S_{km}}{S_{1m}}$. 
Using this modular S-matrix, the particular combination of characters on the 
RHS of \eqref{g2f4chardecomp} is invariant, as it should be.

This, along with the transformation under $T$ which we have already checked, proves \eqref{g2f4chardecomp}.

However, it is clear from the fusion rules, \eqref{g2f4fusion}, 
that something is amiss. If we take the OPE of $[{\bf 7},{\bf 26}]$ with 
itself, the fusion rules dictate that we should see, in addition to the 
desired $[{\bf 1},{\bf 1}]+[{\bf 7},{\bf 26}]$, terms 
involving $[{\bf 7},{\bf 1}]+[{\bf 1},{\bf 26}]$ as well.

While we have managed to reproduce the multiplicity of states correctly, 
it appears that we have failed to reproduce their interactions correctly.  
Moreover Ka\v c and Sanielevici \cite{kacsan} have found several other 
examples of 
non-maximal rank embeddings of characters of affine algebras, of which this 
is, perhaps, the simplest example.  As far as we can tell, the same criticism 
applies to their other examples: the multiplicity of states correctly 
reproduces that of the $E_{8}$ current algebra, but the interactions do not.

It is worth remarking that our previous examples were obtained as (asymmetric) 
orbifolds by some subgroup of the center. 
In the case at hand, $G_{2}$ and $F_{4}$ are center-less\footnote{
This fact is discussed in appendix~\ref{gpthy}.  In addition, they also
have no normal finite subgroup, as any discrete normal subgroup of a connected
group is necessarily central, and there is no center in this case.
The statement on discrete normal subgroups can be shown as follows.
Let $G$ be a connected group and $N$ a discrete normal subgroup.
Let $G$ act on $N$ by conjugation, which it does since $N$ is normal.
Then for any $n \in N$, every $g n g^{-1}$ is in $N$, and connected to
$n$ within $N$, since $G$ is connected.  Since $N$ is discrete,
for $g n g^{-1}$ to be connected to $n$, they must be equal,
hence $N$ is central.  We would like to thank A.~Knutson for pointing
this out to us.
}, so there is 
no obvious orbifold construction that could give rise to \eqref{g2f4chardecomp}.

\section{Symmetric bosonic fibered WZW models}   \label{symmfibwzw}
Now that we have seen alternative constructions of ten-dimensional
heterotic strings using more general current algebras than
$\mathrm{Spin}(16)/\BZ_2$, we will next discuss how to fiber
those current algebras over nontrivial spaces.
As a warm-up, let us first describe a fibered WZW model in the
symmetric case.  This will not be useful for heterotic strings,
but it will provide a good `stepping-stone' to the asymmetric
fibered WZW models we will discuss in the next section.

Start with the total space of a G-bundle in which across coordinate
patches the fibers transform as, $g \mapsto (g_{\alpha \beta}) g
(g_{\alpha \beta}^{-1})$.  Let $A_{\mu}$ be a connection on this bundle.

First, recall from \cite{cliffwzw1}[eqn (2.4)] that a WZW model
in which the adjoint action has been gauged has the form
\begin{equation*}
\begin{split}
S = \: &
- \: \frac{k}{4 \pi} \int_{\Sigma} d^2z \mbox{Tr } \left[
g^{-1} \partial g g^{-1} \overline{\partial} g \right] \\
& - \frac{i k}{12 \pi} \int_B d^3y \epsilon^{ijk}
\mbox{Tr }\left[ g^{-1} \partial_i g g^{-1} \partial_j g 
g^{-1} \partial_k g \right] \\
& + \frac{k}{2 \pi} \int_{\Sigma} d^2z \mbox{Tr }\left[
A_{\overline{z}} g^{-1} \partial g \: - \:
A_{z} \overline{\partial} g g^{-1} \right] \\
& + \frac{k}{2 \pi} \int_{\Sigma} d^2z  
\mbox{Tr } \left[
A_{\overline{z}} g^{-1} A_{z} g \: - \: A_{\overline{z}} A_{z} \right]
\end{split}
\end{equation*}
where $A_z$, $A_{\overline{z}}$ is a worldsheet gauge field.

To define a fibered WZW model, we will want to replace the
worldsheet gauge fields with pullbacks of a gauge field on the 
target space (the connection on the $G$ bundle).
That way, gauge invariance across coordinate patches will be
built in.
Thus,
consider a nonlinear sigma model on the total space of that bundle
with action
\begin{equation*}
\begin{split}
S = \: & \frac{1}{\alpha'} \int_{\Sigma} d^2z \partial_{\alpha} \phi^{\mu}
\partial^{\alpha} \phi^{\nu} g_{\mu \nu}
\: - \: \frac{k}{4 \pi} \int_{\Sigma} d^2z \mbox{Tr } \left[
g^{-1} \partial g g^{-1} \overline{\partial} g \right] \\
& - \frac{i k}{12 \pi} \int_B d^3y \epsilon^{ijk}
\mbox{Tr }\left[ g^{-1} \partial_i g g^{-1} \partial_j g 
g^{-1} \partial_k g \right] \\
& + \frac{k}{2 \pi} \int_{\Sigma} d^2z \mbox{Tr }\left[
\overline{\partial} \phi^{\mu} A_{\mu} g^{-1} \partial g \: - \:
\partial \phi^{\mu} A_{\mu} \overline{\partial} g g^{-1} \right] \\
& + \frac{k}{2 \pi} \int_{\Sigma} d^2z \overline{\partial} 
\phi^{\mu} \partial \phi^{\nu} \mbox{Tr } \left[
A_{\mu} g^{-1} A_{\nu} g \: - \: A_{\mu} A_{\nu} \right]
\end{split}
\end{equation*}
where the $\phi^{\mu}$ are coordinates on the base and $g$ is a coordinate
on the fibers.
On each coordinate patch on the base, the Wess-Zumino term is an
ordinary Wess-Zumino term -- the fields $g$ are fields on the worldsheet,
not functions of the $\phi$ -- and so can be handled in the ordinary
fashion.

Next, although we have deliberately engineered this action to be
well-defined across coordinate patches on the target space,
let us explicitly check that the action is indeed gauge invariant.
Under the following variation
\begin{equation*}
\begin{split}
g \mapsto \: & h g h^{-1} \\
A_{\mu} \mapsto \: & h \partial_{\mu} h^{-1} \: + \:
h A_{\mu} h^{-1}
\end{split}
\end{equation*}
(where $h = h(\phi)$), the variation of all terms except the WZ
term is given by
\begin{multline*}
\delta  = 
 \frac{k}{4 \pi} \int_{\Sigma} d^2z\mbox{Tr }\left[
- h^{-1} \overline{\partial} h g^{-1} \partial g \: + \:
h^{-1} \partial h \overline{\partial} g g^{-1} \: - \:
h^{-1} \partial h g h^{-1} \overline{\partial} h g^{-1} \right. \\
\left. + h^{-1} \partial h g^{-1} \overline{\partial} g
\: - \: \partial g g^{-1} h^{-1} \overline{\partial} h \: + \:
h^{-1} \partial h g^{-1} h^{-1} \overline{\partial} h g \right]
\end{multline*}
and where it is understood that, for example, 
$\partial h = \partial \phi^{\mu} \partial_{\mu} h$.

The variation of the WZ term is given by
\begin{equation*}
\begin{split}
- \frac{3 i k}{12 \pi} &\int_B d^3y \epsilon^{ijk} \mbox{Tr }\left[
g^{-1} h^{-1} \partial_i h h^{-1} \partial_j h \partial_k g \: - \:
g^{-1} h^{-1} \partial_i h h^{-1} \partial_j h g h^{-1} \partial_k h
\right. \\
& + \: h^{-1} \partial_i h \partial_j g g^{-1} \partial_k g g^{-1} \:
- \: g^{-1} h^{-1} \partial_i h \partial_j g h^{-1} \partial_k h \\
& - \: g^{-1} h^{-1} \partial_i h g h^{-1} \partial_j h g^{-1} \partial_k g
\: + \: g^{-1} h^{-1} \partial_i h g h^{-1} \partial_j h h^{-1} \partial_k h\\
& \left. - \: g^{-1} \partial_i g g^{-1} \partial_j g h^{-1} \partial_k h
\: + \: g^{-1} \partial_i g h^{-1} \partial_j h h^{-1} \partial_k h \right] \\
= \: & - \frac{3ik}{12 \pi} \int_B d \mbox{Tr }\left[
- h^{-1} dh \wedge dg g^{-1} \: - \:
h^{-1} dh \wedge g^{-1} dg \: + \: g^{-1} h^{-1} (dh) g \wedge h^{-1} dh
\right] \\
= \: & - \frac{3ik}{12\pi}\int_{\Sigma} \mbox{Tr }\left[
- h^{-1} dh \wedge dg g^{-1} \: - \:
h^{-1} dh \wedge g^{-1} dg \: + \: g^{-1} h^{-1} (dh) g \wedge h^{-1} dh
\right] 
\end{split}
\end{equation*}

If we write $z = x + iy$ then
\begin{equation*}
dz \wedge d\overline{z} \: - \: d \overline{z} \wedge dz \: = \:
2 i \left( dy \wedge dx \: - \: dx \wedge dy \right)
\end{equation*}
then we see that the terms generated by the variation of the WZ term
are exactly what is needed to cancel the terms generated by everything else.

Note that the computation above, the check that the model is
well-defined across target-space coordinate patches,
is identical to the computation needed to show that an ordinary
gauged WZW model is invariant under gauge transformations.

The model we have described so far is bosonic, but one could
imagine adding fermions along the base and demanding supersymmetry
under transformations that leave the fibers invariant.
A simpler version of this is obtained by taking a $(2,2)$ nonlinear
sigma model and adding right- and left-moving fermions $\lambda_{\pm}$
coupling to a vector bundle over the $(2,2)$ base.
Demanding that the resulting model be $(2,2)$
supersymmetric on-shell unfortunately forces the bundle to be flat:  $F=0$.
Roughly, half of the constraints one obtains from supersymmetry
force the curvature
to be holomorphic, in the sense $F_{ij} = 
F_{\overline{\imath} \overline{\jmath}} = 0$,
and the other half force the connection to be flat.
We shall find in the next section that imposing merely $(0,2)$
supersymmetry is easier:  one merely needs the curvature to
be holomorphic, not necessarily flat.

\section{Fibered (0,2) WZW models}   \label{chirfibwzw}

\subsection{Construction of the lagrangian}
Begin with some principal $G$ bundle with connection $A_{\mu}$
over some Calabi-Yau $X$.
Consider a nonlinear sigma model on the total space of that bundle.
We shall think of the fibers as defining, locally, WZW models,
so we use the connection $A_{\mu}$ to define a chiral multiplication on the
fibers of the bundle, and have a WZ term to describe $H$ flux in the fibers.

\subsubsection{Gauge invariance and global well-definedness}
We are going to write down a fibered WZW model in which each
fiber is a gauged WZW model, gauging the action $g \mapsto h g$
across coordinate patches on the target space, the principal
$G$ bundle.

First, recall from \cite{cliffwzw1}[eqn (2.9)]
and \cite{wittenholfac}, a gauged WZW model
gauging the chiral multiplication $g \mapsto h g$ is given by
\begin{equation*}
\begin{split}
S' = \: &
- \: \frac{k}{4 \pi} \int_{\Sigma} d^2 z \mbox{Tr }\left(
g^{-1} \partial_z g g^{-1} \partial_{\overline{z}} g \right)
\: - \: \frac{ik}{12 \pi} \int_B d^3 y \epsilon^{ijk}
\mbox{Tr }\left( g^{-1} \partial_i g g^{-1} \partial_j g
g^{-1} \partial_k g \right) \\
& - \: \frac{k}{2 \pi} \int_{\Sigma} d^2 z \mbox{Tr }
\left( A_{z} 
\partial_{\overline{z}} g g^{-1}\: + \: \frac{1}{2}  
A_{z} A_{\overline{z}} \right)
\end{split}
\end{equation*}
where $A_z$, $A_{\overline{z}}$ are worldsheet gauge fields.

With that in mind, to describe a fibered WZW model, one would
replace the worldsheet gauge fields with pullbacks of a connection
$A_{\mu}$ on the target space, the principal $G$ bundle.
In fact,  
one would initially suppose that the action should have the form
\begin{equation*}
\begin{split}
S = \: & \frac{1}{\alpha'} \int_{\Sigma} d^2z\left( \frac{1}{4}
g_{i \overline{\jmath}}
\partial_{\alpha} \phi^{i} \partial^{\alpha} \phi^{\overline{\jmath}}
\: + \: i g_{i \overline{\jmath}} \psi_+^{\overline{\jmath}} D_{\overline{z}}
\psi_+^i \right) \\
& - \: \frac{k}{4 \pi} \int_{\Sigma} d^2 z \mbox{Tr }\left(
g^{-1} \partial_z g g^{-1} \partial_{\overline{z}} g \right)
\: - \: \frac{ik}{12 \pi} \int_B d^3 y \epsilon^{ijk}
\mbox{Tr }\left( g^{-1} \partial_i g g^{-1} \partial_j g
g^{-1} \partial_k g \right) \\
& - \: \frac{k}{2 \pi} \int_{\Sigma} d^2 z \mbox{Tr }
\left( (\partial_{z} \phi^{\mu}) A_{\mu} 
\partial_{\overline{z}} g g^{-1}\: + \: \frac{1}{2} (\partial_z \phi^{\mu} 
\partial_{\overline{z}} \phi^{\nu}) A_{\mu} A_{\nu} \right)
\end{split}
\end{equation*}
The field $g$ defines a coordinate on the fibers of the bundle,
and $\phi$ are coordinates on the base.

However, the full analysis is slightly more complicated.
As described in \cite{cliffwzw1,wittenholfac,ralph1} a WZW action is not
invariant under chiral group multiplications, so the action above is
not invariant across coordinate patches on the target space.
Specifically, under the target-space gauge transformation
\begin{equation*}
\begin{split}
g \mapsto \: & h g \\
A_{\mu} \mapsto \: & h A_{\mu} h^{-1} \: + \: h \partial_{\mu} h^{-1}
\end{split}
\end{equation*}
(where $h$ is a group-valued function on the target space)
the gauge transformation of the terms above excepting the Wess-Zumino term
is given by
\begin{equation*}
\frac{k}{4 \pi} \int_{\Sigma} d^2z \mbox{Tr }\left( h^{-1} \partial h
\overline{\partial} g g^{-1} \: - \: h^{-1} \overline{\partial} h \partial g
g^{-1} \: + \: \overline{\partial} \phi^{\mu} A_{\mu} h^{-1} \partial h
\: - \: \partial \phi^{\mu} A_{\mu} h^{-1} \overline{\partial} h \right)
\end{equation*}
where, for example, $\partial h = (\partial_z \phi^{\mu})( \partial_{\mu} h)$,
and the gauge transformation of the Wess-Zumino term is given by
\begin{equation*}
- \frac{ik}{12 \pi} \int_B d^3y \epsilon^{ijk} \mbox{Tr }\left(
h^{-1} \partial_i h h^{-1} \partial_j h h^{-1} \partial_k h \right)
\: + \: \frac{ik}{4 \pi} \int_{\Sigma} \mbox{Tr }\left(
h^{-1} dh \wedge dg g^{-1} \right)
\end{equation*}

This lack of gauge invariance is exactly what one would expect
of a bosonized description of the left-movers on a heterotic
string worldsheet.  There is a chiral gauge anomaly in the
fermionic realization which after bosonization should be realized classically.
On the other hand, a lack of gauge-invariance across coordinate
patches means we have a problem with global well-definedness of the chiral fibered WZW
model.

We can resolve this problem with gauge invariance in the standard
way for heterotic strings:  assign the $B$ field nontrivial
gauge transformation properties.  So, we add a $B$ field,
coupling as
\begin{equation*}
\frac{1}{\alpha'} \int_{\Sigma} d^2z B_{\mu \nu} \left(
\partial \phi^{\mu} \overline{\partial} \phi^{\nu} \: - \:
\overline{\partial} \phi^{\mu} \partial \phi^{\nu} \right)
\end{equation*}
and demand that
under the gauge transformation above, the holonomy above pick up
the terms 
\begin{equation}  \label{CS-gauge-trans}
+ \frac{ik}{12 \pi} \int_B d^3y \epsilon^{ijk} \mbox{Tr }\left(
h^{-1} \partial_i h h^{-1} \partial_j h h^{-1} \partial_k h \right) \: + \:
\frac{ik}{4 \pi} \int_{\Sigma} \mbox{Tr }\left(
h^{-1} dh \wedge d \phi^{\mu} A_{\mu} \right)
\end{equation}
This transformation law manifestly restores gauge-invariance.

Let us check for a minute that this transformation law is consistent.
The second term is a two-form, and so it is completely consistent
for the $B$ field to pick up such a term.  The first term, on the
other hand, is a three-form, which in general will not even
be closed on each overlap chart.  As a result, the first term cannot
be expressed even locally in terms of a two-form.

However, there is a fix.
In addition to gauge invariance, we must also demand, as is standard
in heterotic strings, that the $B$ field transform under local
Lorentz transformations acting on the chiral right-moving fermions.  
These transformations are anomalous, and by demanding that the $B$
field transform, we can restore the gauge-invariance broken by
the anomalies.
Under such transformations, the $B$ field
will necessarily pick up two closely analogous terms, one of which
will involve another problematic three-form.
Thus, we need for the combination
\begin{equation*}
k \, \mbox{Tr } \left( \left( g_{\alpha \beta}^F \right)^{-1} d g_{\alpha \beta}^F
\right)^3 \: - \:
\mbox{Tr }\left( \left( g_{\alpha \beta}^R \right)^{-1} d g_{\alpha \beta}^R
\right)^3 
\end{equation*}
to be exact on each overlap, where the $g_{\alpha \beta}$'s are transition
functions for the gauge ($F$) and tangent ($R$) bundles.
This turns out \cite{tonypriv} to be implied by the statement that
$k \mbox{Tr }F^2$ and $\mbox{Tr } R^2$ match in cohomology;
writing Chern-Simons forms for both and interpreting in terms of
Deligne cohomology, the condition that the difference across overlaps
is exact is immediate.
This is the first appearance of the anomaly-cancellation constraint that
\begin{equation}    \label{anom1}
k \, [ \mbox{Tr } F^2 ] \: = \: [ \mbox{Tr } R^2 ]
\end{equation}
where $k$ is the level of the fibered Kac-Moody algebra.
We shall see this same constraint emerge several more times in
different ways.

In any event, so long as the condition~(\ref{anom1}) is obeyed,
we see that the chiral fibered WZW model is well-defined globally.
Next we shall the fermion kinetic terms in this model.

In order to formulate a supersymmetric theory, we shall need to
add a three-form flux $H_{\mu \nu \rho}$ to the connection appearing
in the $\psi$ kinetic terms.  Ordinarily $H = d B$, but we need
$H$ to be gauge- and local-Lorentz-neutral, whereas $B$ transforms
under both gauge and local Lorentz transformations.  To fix this,
we follow the standard procedure in heterotic strings of adding
Chern-Simons terms.  For example, the gauge terms~(\ref{CS-gauge-trans})
are the same as those arising in a gauge transformation of the
Chern-Simons term
\begin{equation*}
 + \: \frac{ i k }{ 4 \pi} \int_B d^3y \epsilon^{ijk}
\partial_i \phi^{\mu} \partial_j \phi^{\nu} \partial_k \phi^{\rho}
\mbox{Tr }\left( A_{\mu} \partial_{\nu} A_{\rho} \: + \:
\frac{2}{3}A_{\mu}A_{\nu}A_{\rho} \right) 
\end{equation*}
and similarly one can cancel the terms picked up under
local Lorentz transformation by adding a term involving the
Chern-Simons form coupling to the spin connection.
Schematically, we have
\begin{equation*}
H \: = \: d B \: + \: (\alpha')\mbox{Tr }\left( k \, CS(A) \: - \: CS(\omega) \right)
\end{equation*}
where $k$ is the level of the fibered current algebra.
$H$ is now an ordinary gauge- and local-Lorentz-invariant three-form.
This statement implies that $k \, \mbox{Tr } F^2$ and $\mbox{Tr }R^2$
must be in the same cohomology class.  For a fibered current algebra
defined by a principal $SU(n)$ bundle ${\cal E}$ over a space $X$,
this is the statement that $k \, c_2({\cal E}) = c_2(TX)$,
which generalizes the ordinary anomaly cancellation condition of heterotic
strings.  This is the second appearance of this constraint;
we shall see it again later.

As an aside, note that since this model has
nonzero $H$ flux, the metric cannot be K\"ahler
\cite{strominger}.
More precisely, to zeroth order in $\alpha'$ a K\"ahler metric can
be consistent, but to next leading order in $\alpha'$ the metric
will be nonK\"ahler, with $H$ measuring how far the metric is from
being K\"ahler.

Also note that this analysis is analogous to, though slightly
different from, that of $(0,2)$ WZW models discussed in
\cite{cliffwzw1,ralph1}.  There, WZW models with chiral group
multiplications and chiral fermions were also considered.  However,
the fermions lived in the tangent bundle to the group manifold,
so the chiral group multiplication induced the right-moving fermion
anomaly, and so that chiral fermion anomaly and the classical
noninvariance of the action could be set to cancel each other out.
Here, on the other hand, the chiral fermions live on the base,
not the WZW fibers, and so do not see the chiral group multiplication
(which only happens on the fibers).  Thus, here we proceed in a more
nearly traditional fashion, by adding a $B$ field with nontrivial
gauge- and local-Lorentz transformations, whose global well-definedness
places constraints on the bundles involved.

Thus, the gauge-invariant action has the form
\begin{equation*}
\begin{split}
S = \: & \frac{1}{\alpha'} \int_{\Sigma} d^2z\left( \frac{1}{4}
g_{i \overline{\jmath}}
\partial_{\alpha} \phi^{i} \partial^{\alpha} \phi^{\overline{\jmath}}
\: + \: i g_{i \overline{\jmath}} \psi_+^{\overline{\jmath}} D_{\overline{z}}
\psi_+^i \right) \\
& + \: \frac{1}{\alpha'} \int_{\Sigma} d^2z B_{\mu \nu}
\left(
\partial \phi^{\mu} \overline{\partial} \phi^{\nu} \: - \:
\overline{\partial} \phi^{\mu} \partial \phi^{\nu} \right) \\
& - \: \frac{k}{4 \pi} \int_{\Sigma} d^2 z \mbox{Tr }\left(
g^{-1} \partial_z g g^{-1} \partial_{\overline{z}} g \right)
\: - \: \frac{ik}{12 \pi} \int_B d^3 y \epsilon^{ijk}
\mbox{Tr }\left( g^{-1} \partial_i g g^{-1} \partial_j g
g^{-1} \partial_k g \right) \\
& - \: \frac{k}{2 \pi} \int_{\Sigma} d^2 z \mbox{Tr }
\left( (\partial_{z} \phi^{\mu}) A_{\mu} 
\partial_{\overline{z}} g g^{-1}\: + \: \frac{1}{2} (\partial_z \phi^{\mu} 
\partial_{\overline{z}} \phi^{\nu}) A_{\mu} A_{\nu} \right) \\
\end{split}
\end{equation*}

\subsubsection{Worldsheet supersymmetry}
Next, let us demand that the model possess $(0,2)$ supersymmetry,
under the transformations
\begin{equation*}
\begin{split}
\delta \phi^i = \: & i \alpha_- \psi_+^i \\
\delta \phi^{\overline{\imath}} = \: & i \tilde{\alpha}_- \psi_+^{
\overline{\imath}} \\
\delta \psi_+^i = \: & - \tilde{\alpha}_- \partial \phi^i \\
\delta \psi_+^{\overline{\imath}} = \: & - \alpha_- \partial
\phi^{\overline{\imath}} \\
\delta g = \: & 0
\end{split}
\end{equation*}
Supersymmetry will require us to add the gauge-invariant term
\begin{equation*}
\frac{i k}{4 \pi} \int_{\Sigma} d^2z \mbox{Tr }\left(
F_{\mu \nu}  \overline{\partial}_A g g^{-1} \right)
\psi_+^{\mu} \psi_+^{\nu}
\end{equation*}
where
\begin{equation*}
\begin{split}
\overline{\partial}_A g g^{-1} = \: & \left( \overline{\partial} g 
\: + \: \overline{\partial} \phi^{\mu} A_{\mu} g \right) g^{-1} \\
= \: & \overline{\partial} g g^{-1} \: + \: \overline{\partial} \phi^{\mu} 
A_{\mu}
\end{split}
\end{equation*}
and $F_{\mu \nu} = \partial_{\mu} A_{\nu} - \partial_{\nu} A_{\mu}
+ [A_{\mu}, A_{\nu}]$.
The term above is an analogue of the four-fermi term appearing
in standard heterotic string constructions.
We shall also add an $H$ flux field to the base.
One finds that for the supersymmetry transformations to close,
one needs $F_{ij} = F_{\overline{\imath} \overline{\jmath}} = 0$.

Let us outline how the $\alpha_-$ supersymmetry transformations work.

The $\alpha_-$ terms in the supersymmetry transformation of the base terms
\begin{equation*}
 \frac{1}{\alpha'} \int_{\Sigma} d^2z\left( \frac{1}{4}
g_{i \overline{\jmath}}
\partial_{\alpha} \phi^{i} \partial^{\alpha} \phi^{\overline{\jmath}}
\: + \: \frac{i}{2} g_{\mu \nu} \psi_+^{\mu} D_{\overline{z}}
\psi_+^{\nu} 
\: + \: B_{\mu \nu} \left(
\partial \phi^{\mu} \overline{\partial} \phi^{\nu} 
\: - \: \overline{\partial} \phi^{\mu} \partial \phi^{\nu}
\right) \right) 
\end{equation*}
where
\begin{equation*}
D_{\overline{z}} \psi_+^{\nu} \: = \: \overline{\partial} \psi_+^{\nu} \: + \:
\overline{\partial} \phi^{\mu} \left( \Gamma^{\nu}_{\: \: \sigma \mu}
\: - \: H^{\nu}_{\: \: \sigma \mu} \right) \psi_+^{\sigma}
\end{equation*}
are given by
\begin{equation*}
\begin{split}
\frac{1}{\alpha'}\int d^2z & \left[ 
(i \alpha_- \psi_+^i) (\overline{\partial} \phi^{\mu})
(\partial \phi^{\nu}) (H_{i \mu \nu}) \right] \\
+& \frac{1}{\alpha'} \int_{\Sigma} d^2z 
\left[ \frac{ i}{2} (i \alpha_- \psi_+^i )(\overline{\partial} \phi^{\mu})
\psi_+^j \psi_+^{\overline{k}} 
\left( H_{\overline{k} i j, \mu} \: - \: H_{\overline{k} i \mu, j} \: - \:
H_{j \overline{k} \mu, i} \: + \: H_{j i \mu, \overline{k}} \right)
\right] \\
+& \frac{1}{\alpha'} \int_{\Sigma} d^2z (i \alpha_- \psi_+^i)
\left( B_{\mu \nu, i} \: - \: B_{i \nu, \mu} \: - \: B_{\mu i, \nu}
\right) \left( \partial \phi^{\mu} \overline{\partial} \phi^{\nu}
\: - \: \overline{\partial} \phi^{\mu} \partial \phi^{\nu} \right)
\end{split}
\end{equation*}
and where we needed to assume 
\begin{equation*}
\begin{array}{c}
H_{ijk} \: = \: 
H_{\overline{\imath} \overline{\jmath} \overline{k}} \: = \: 0 \\
H_{i j \overline{k}} \: = \: \frac{1}{2}\left( g_{i \overline{k}, j}
\: - \: g_{j \overline{k}, i} \right) \: = \: \Gamma_{i j \overline{k}}
\end{array}
\end{equation*}
(This was derived off-shell, without using any equations of motion.)

The $\alpha_-$ terms in the supersymmetry transformation of the fiber terms
\begin{equation*}
\begin{split}
- \: \frac{k}{4 \pi} \int_{\Sigma} d^2 z & \mbox{Tr }\left(
g^{-1} \partial_z g g^{-1} \partial_{\overline{z}} g \right)
\: - \: \frac{ik}{12 \pi} \int_B d^3 y \epsilon^{ijk}
\mbox{Tr }\left( g^{-1} \partial_i g g^{-1} \partial_j g
g^{-1} \partial_k g \right) \\
&  - \: \frac{k}{2 \pi} \int_{\Sigma} d^2 z \mbox{Tr }
\left( (\partial_{z} \phi^{\mu}) A_{\mu} 
\partial_{\overline{z}} g g^{-1}\: + \: \frac{1}{2} (\partial_z \phi^{\mu} 
\partial_{\overline{z}} \phi^{\nu}) A_{\mu} A_{\nu} \right)  \\
& + \: \frac{i k}{4 \pi} \int_{\Sigma} d^2z \mbox{Tr }\left(
F_{\mu \nu} \overline{\partial}_A g g^{-1} \right)
\psi_+^{\mu} \psi_+^{\nu}
\end{split}
\end{equation*}
are given by
\begin{multline*}
\frac{i k}{4 \pi} \int_{\Sigma} d^2z 
\left( i \alpha_- \psi_+^i \right)
\mbox{Tr }\left( F_{\mu \nu} F_{i \lambda} \right)
\psi_+^{\mu} \psi_+^{\nu} 
(\overline{\partial} \phi^{\lambda}) \\
\: - \: \frac{k}{4 \pi} \int_{\Sigma} d^2z 
\left( i \alpha_- \psi_+^i \right)
\overline{\partial} \phi^{\mu} \partial \phi^{\nu} \mbox{Tr }\left(
\left( A_i \partial_{\mu} A_{\nu} \: + \: \frac{2}{3} A_i A_{\mu} A_{\nu}
\right) \: \pm \: \mbox{ permutations }
\right)
\end{multline*}
The supersymmetry transformations only close on-shell\footnote{Alternatively,
the supersymmetry transformations will close off-shell if instead of
$\delta g = 0$ we take
\begin{equation*}
\delta g \: = \: - (i \alpha_- \psi_+^i) A_i g \: - \: (i \tilde{\alpha}_-
\psi_+^{\overline{\imath}}) A_{\overline{\imath}} g
\end{equation*}
(This is true for both $\alpha_-$ transformations considered here
as well as $\tilde{\alpha}_-$ transformations.)
In this form supersymmetry transformations explicitly commute with
gauge transformations; on the other hand, the on-shell formulation
$\delta g = 0$ makes it explicit that supersymmetry is only
meaningfully acting on the base.};
to get the result above requires using the classical
equations of motion for $g$,
namely
\begin{equation}  \label{2ndclassconstr}
\partial_A \left( \overline{\partial}_A g g^{-1} \right) \: = \:
\partial \phi^{\mu} \overline{\partial} \phi^{\nu} F_{\mu \nu} 
\: + \: \frac{i}{2} [ F_{\mu \nu}, \overline{\partial}_A g g^{-1}] 
\psi_+^{\mu} \psi_+^{\nu} \: + \:
\frac{i}{2} \overline{\partial}_A\left( F_{\mu \nu} \psi_+^{\mu}
\psi_+^{\nu} \right)
\end{equation}
where
\begin{equation*}
\partial_A\left( \overline{\partial}_A g g^{-1} \right) \: = \:
\partial \left( \overline{\partial}_A g g^{-1} \right) \: + \:
[ \partial \phi^{\lambda} A_{\lambda}, \overline{\partial}_A g g^{-1} ]
\end{equation*}
Note equation~(\ref{2ndclassconstr})  generalizes the chirality condition
$\partial( \overline{\partial} g g^{-1} ) = 0$ that appears in
ordinary (non-fibered) WZW models.

We will also use equation~(\ref{2ndclassconstr}) to define a second
class constraint -- we are describing chiral nonabelian bosons,
after all.

Also note equation~(\ref{2ndclassconstr}) is the supersymmetrization 
of the anomaly in the chiral gauge current:
defining $j = \overline{\partial}_A g g^{-1}$, and omitting fermions,
this says $D j \propto F$.  If the WZW current were realized by fermions,
this would be the chiral anomaly; here, we have bosonized, and so the
anomaly is realized classically.
In such a fermionic realization, the second term is a classical
contribution to the divergence of the current from the four-fermi term
in the action, and the third term is a non-universal contribution to the
anomaly from a one-loop diagram also involving the four-fermi interaction.

In a fermionic realization of the left-movers, the terms in the
supersymmetry transformations above would
not appear at zeroth order in $\alpha'$.  Classically, supersymmetry
transformations of the action result in one-fermi terms proportional
to $H - dB$ and three-fermi terms proportional to $dH$, both of
which are proportional to $\alpha'$.
However, at next-to-leading-order in $\alpha'$ on the worldsheet,
one has more interesting effects.  Specifically, 
``supersymmetry anomalies'' arise \cite{sen1,sen2}.
These are phase factors picked up by the path integral measure.
Unlike true anomalies, these are cancelled by counterterms.
In particular, the Chern-Simons terms added to make $H$ gauge- and
local-Lorentz-invariant cancel out the effect of these `anomalies.'

In more detail, if we realize the left-moving gauge degrees of
freedom by chiral fermions $\lambda_-$,
we can realize worldsheet supersymmetry off-shell\footnote{If we take
$\delta \lambda_- = 0$, the worldsheet supersymmetry transformations close
only if one uses the $\lambda_-$ equations of motion.} with supersymmetry
transformations of the form
\begin{equation*}
\delta \lambda_- \: = \: - (i \alpha_- \psi_+^i) A_i \lambda_-
\: - \: (i \tilde{\alpha}_- \psi_+^{\overline{\imath}}) A_{\overline{\imath}}
\lambda_-
\end{equation*}
where $A_{\mu}$ is the target-space gauge field.
However, these supersymmetry transformations are equivalent to
(anomalous chiral) gauge transformations with parameter
\begin{equation}    \label{susygauge}
- (i \alpha_- \psi_+^i) A_i \: - \: 
(i \tilde{\alpha}_- \psi_+^{\overline{\imath}}
) A_{\overline{\imath}}
\end{equation}
Thus, the supersymmetry transformation implies an anomalous gauge
transformation, and so the path integral measure picks up a phase factor.
From the (universal) bosonic term in the divergence of the 
gauge current proportional to the
curvature $F$, we will get a one-fermi term in the anomalous
transformation proportional to the Chern-Simons form.
In our case, as we have bosonized the left-movers, we get such
a one-fermi term in supersymmetry transformations classically.
In addition to the universal piece, there is a regularization-dependent
multifermi contribution as well.
If we calculate the anomalous divergence of the gauge current
in a fermionic realization, then because of the four-fermi term
$F \lambda \lambda \psi \psi$ there will be a two-fermi contribution
to the divergence of the gauge current proportional to 
$\overline{\partial} (F \psi \psi)$.  Plugging into the
gauge parameter~(\ref{susygauge}) yields a three-fermi term
in the supersymmetry transformations proportional to
$\mbox{Tr } F\wedge F$, exactly as we have discovered in the
classical supersymmetry transformations of our bosonized
formulation.

There is a closely analogous phenomenon of supersymmetry anomalies
in the right-moving fermions as well.  Since we have not bosonized
them, the analysis here is identical to that for ordinary
heterotic string constructions discussed, for example, in
\cite{sen1,sen2}.  In terms of supersymmetry transformations
of the right-moving fermions written with general-covariant
indices, {\it e.g.} $\delta \psi_+^i \: = \: - \tilde{\alpha}_-
\partial \phi^i$, the source of the anomaly is not obvious.
To make it more manifest, we must switch to local Lorentz indices,
and define
\begin{equation*}
\psi_+^a \: = \: e^a_{\mu} \psi_+^{\mu}
\end{equation*}
Then, the supersymmetry transformations have the form
\begin{equation*}
\delta \psi_+^a \: = \: \left( e^a_i (- \tilde{\alpha}_- \partial
\phi^i) \: + \: e^a_{\overline{\imath}}(- \alpha_- \partial \phi^{
\overline{\imath}}) \right) \: + \:
\left( e^a_{\mu,i} (i \alpha_- \psi_+^i) \psi_+^{\mu} \: + \:
e^a_{\mu,\overline{\imath}} (i \tilde{\alpha}_- \psi_+^{\overline{\imath}})
\psi_+^{\mu} \right)
\end{equation*}
The second set of terms above can be written as
(anomalous, chiral) local Lorentz
transformations, and so the supersymmetry transformations induce
anomalous local Lorentz transformations.
In particular, under a supersymmetry transformation the path integral
measure will pick up a phase factor including a one-fermi term
proportional to the Chern-Simons form for the target-space spin
connection, whose origin is the (univeral, bosonic) curvature term
in the divergence of the local Lorentz current.
The path integral phase factor will also include a multifermi contribution.
Here, the same analysis of four-fermi terms as before would appear
to imply that the multifermi contribution will be proportional to
$FR$, where $F$ is the gauge curvature and $R$ is the metric curvature.
However, these multifermi terms are sensitive to the choice of regulator,
and to maintain (0,2) worldsheet supersymmetry we must be very careful
about the choice of regulator here.  For the correct choice of regularization,
the multifermi contribution is a three-fermi term proportional to
$\mbox{Tr } R \wedge R$, where $R$ is the curvature of the
connection $\Gamma - H$, as discussed in {\it e.g.} \cite{sen1,sen2}.

As a check on this method, note that if we replace the right-moving
chiral fermions with nonabelian bosons, then following the same analysis
as for the gauge degrees of freedom the supersymmetry transformations 
will automatically generate one-fermi and three-fermi terms of the
desired form.

For more information on supersymmetric anomalies in such two-dimensional
theories, see also \cite{wangwu,ssw}.  See also \cite{yuwu1,yuwu2}
for an interesting approach to the interaction of second-class
constraints and worldsheet supersymmetry.

To summarize, under (anomalous) worldsheet supersymmetry transformations we have
found one-fermi terms proportional to 
\begin{equation*}
H \: - \: d B \: - \: (\alpha')\left ( k \, \mbox{CS}(A) \: - \: \mbox{CS}(\omega - H)
\right)
\end{equation*}
and three-fermi terms proportional to 
\begin{equation*}
dH \: - \: (\alpha')\left( k \, \mbox{Tr } F \wedge F \: - \:
\mbox{Tr } R \wedge R \right)
\end{equation*}
where the terms involving the spin connection $\omega$ arise
from quantum corrections, and the terms involving the gauge field
$A$ arise classically in our bosonic construction but from
quantum corrections in fermionic realizations of left-movers.
Closure of supersymmetry is guaranteed by our definition of $H$.
Put another way, we see that worldsheet supersymmetry is deeply
intertwined with the Green-Schwarz mechanism.

The $\tilde{\alpha}_-$ terms in the supersymmetry transformations are
almost identical.  
The $\tilde{\alpha}_-$ terms in the supersymmetry transformation
of the base terms are given by
\begin{equation*}
\begin{split}
\frac{1}{\alpha'} \int_{\Sigma} d^2z &(i \tilde{\alpha}_- \psi_+^{\overline{k}})
(\overline{\partial} \phi^{\mu}) (\partial \phi^{\nu}) H_{
\overline{k} \mu \nu } \\
& + \: \frac{1}{\alpha'} \int_{\Sigma} d^2z \frac{i}{2}
(i \tilde{\alpha}_- \psi_+^{\overline{k}}) (\overline{\partial} \phi^{\mu})
\psi_+^i \psi_+^{\overline{\jmath}}\left(
H_{\overline{\jmath} \overline{k} i, \mu} \: - \:
H_{\overline{\jmath} \overline{k} \mu, i} \: - \:
H_{i \overline{\jmath} \mu, \overline{k}} \: + \:
H_{i \overline{k} \mu, \overline{\jmath}} \right) \\
& + \: \frac{1}{\alpha'} \int_{\Sigma} d^2z ( i \tilde{\alpha}_-
\psi_+^{\overline{\imath}} ) \left(
B_{\mu \nu, \overline{\imath}} \: - \: B_{\overline{\imath} \nu, \mu} \: - \:
B_{\mu \overline{\imath}, \nu} \right)\left(
\partial \phi^{\mu} \overline{\partial} \phi^{\nu} \: - \:
\overline{\partial} \phi^{\mu} \partial \phi^{\nu} \right)
\end{split}
\end{equation*}
which are virtually identical to the corresponding $\alpha_-$ terms.

The $\tilde{\alpha}_-$ terms in the supersymmetry transformation of the
fiber terms are given by
\begin{equation*}
\begin{split}
\frac{i k}{4 \pi} \int_{\Sigma} d^2z & \mbox{Tr }
\left( F_{\mu \nu} F_{\overline{\imath}
\rho} \right) (i \tilde{\alpha}_- \psi_+^{\overline{\imath}} )
\psi_+^{\mu} \psi_+^{\nu} \overline{\partial} \phi^{\rho}\\
& - \: \frac{k}{4 \pi} \int_{\Sigma} d^2 z
(i \tilde{\alpha}_- \psi_+^{\overline{\imath}}) (\overline{\partial} \phi^{\mu})
(\partial \phi^{\nu}) \mbox{Tr }\left( A_{\overline{\imath}} \partial_{\mu}
A_{\nu} \: + \: \frac{2}{3} A_{\overline{\imath}} A_{\mu} A_{\nu}
\: \pm \: \mbox{permutations} \right)
\end{split}
\end{equation*}
which are virtually identical to the corresponding $\alpha_-$ terms.
(As before, to get the result above requires using the equations of
motion for $g$.)

The supersymmetry anomaly story works here in the same way as for
the $\alpha_-$ terms, and just as for the $\alpha_-$ terms,
one can show that the worldsheet theory is supersymmetric through
first order in $\alpha'$.

\subsubsection{The full gauge-invariant supersymmetric lagrangian}
Let us summarize the results of the last two subsections.
The full lagrangian is given by
\begin{equation*}
\begin{split}
S = \: & \frac{1}{\alpha'} \int_{\Sigma} d^2z\left( \frac{1}{4}
g_{i \overline{\jmath}}
\partial_{\alpha} \phi^{i} \partial^{\alpha} \phi^{\overline{\jmath}}
\: + \: 
\frac{i}{2} g_{\mu \nu} \psi_+^{\mu} D_{\overline{z}}
\psi_+^{\nu}
\right) \\
& + \: \frac{1}{\alpha'} \int_{\Sigma} d^2z B_{\mu \nu} \left(
\partial \phi^{\mu} \overline{\partial} \phi^{\nu} \: - \:
\overline{\partial} \phi^{\mu} \partial \phi^{\nu} \right) \\
& - \: \frac{k}{4 \pi} \int_{\Sigma} d^2 z \mbox{Tr }\left(
g^{-1} \partial_z g g^{-1} \partial_{\overline{z}} g \right)
\: - \: \frac{ik}{12 \pi} \int_B d^3 y \epsilon^{ijk}
\mbox{Tr }\left( g^{-1} \partial_i g g^{-1} \partial_j g
g^{-1} \partial_k g \right) \\
& - \: \frac{k}{2 \pi} \int_{\Sigma} d^2 z \mbox{Tr }
\left( (\partial_{z} \phi^{\mu}) A_{\mu} 
\partial_{\overline{z}} g g^{-1}\: + \: \frac{1}{2} (\partial_z \phi^{\mu} 
\partial_{\overline{z}} \phi^{\nu}) A_{\mu} A_{\nu} \right) \\
& + \: \frac{i k}{4 \pi} \int_{\Sigma} d^2z \mbox{Tr }\left(
F_{\mu \nu} \overline{\partial}_A g g^{-1} \right) 
\psi_+^{\mu} \psi_+^{\nu}
\end{split}
\end{equation*}
where
\begin{equation*}
D_{\overline{z}} \psi_+^{\nu} \: = \: \overline{\partial} \psi_+^{\nu} \: + \:
\overline{\partial} \phi^{\mu} \left( \Gamma^{\nu}_{\: \: \sigma \mu}
\: - \: H^{\nu}_{\: \: \sigma \mu} \right) \psi_+^{\sigma}
\end{equation*}
and the metric $g_{\mu \nu}$ on the base will not be K\"ahler
(except optionally at zeroth order in $\alpha'$).

The action is well-defined under the gauge transformations 
\begin{equation*}
\begin{split}
g \mapsto \: & h g \\
A_{\mu} \mapsto \: & h A_{\mu} h^{-1} \: + \: h \partial_{\mu} h^{-1}
\end{split}
\end{equation*}
across coordinate-charge-changes on the base,
where $h$ is a group-valued function on the overlap patch on the target space,
and the $B$ field transforms to absorb both the gauge anomaly above
and the local Lorentz anomaly on the right-moving chiral fermions.

The action is also invariant under the (0,2) worldsheet supersymmetry
transformations
\begin{equation*}
\begin{split}
\delta \phi^i = \: & i \alpha_- \psi_+^i \\
\delta \phi^{\overline{\imath}} = \: & i \tilde{\alpha}_- \psi_+^{
\overline{\imath}} \\
\delta \psi_+^i = \: & - \tilde{\alpha}_- \partial \phi^i \\
\delta \psi_+^{\overline{\imath}} = \: & - \alpha_- \partial
\phi^{\overline{\imath}} \\
\delta g = \: & 0
\end{split}
\end{equation*}
where we assume $F_{ij} = F_{\overline{\imath} \overline{\jmath}} = 0$,
and that $H$ has only (1,2) or (2,1) components, no (0,3) or (3,0),
related to the metric by
\begin{equation*}
H_{i \overline{\jmath} k} \: = \: - \frac{1}{2}\left(
g_{i \overline{\jmath}, k} \: - \: g_{k \overline{\jmath}, i} \right)
\end{equation*}
and where $H$ is also given by the difference of Chern-Simons forms,
in the form
\begin{equation*}
H \: = \: dB \: + \: (\alpha')\left(k \, \mbox{CS}(A) \: - \:
\mbox{CS}(\omega - H)\right)
\end{equation*}

The classical equations of motion for $g$ are
\begin{equation*}
\partial_A \left( \overline{\partial}_A g g^{-1} \right) \: = \:
\partial \phi^{\mu} \overline{\partial} \phi^{\nu} F_{\mu \nu} 
\: + \: \frac{i}{2} [ F_{\mu \nu}, \overline{\partial}_A g g^{-1}] 
\psi_+^{\mu} \psi_+^{\nu} \: + \:
\frac{i}{2} \overline{\partial}_A\left( F_{\mu \nu} \psi_+^{\mu}
\psi_+^{\nu} \right)
\end{equation*}
where
\begin{equation*}
\partial_A\left( \overline{\partial}_A g g^{-1} \right) \: = \:
\partial \left( \overline{\partial}_A g g^{-1} \right) \: + \:
[ \partial \phi^{\lambda} A_{\lambda}, \overline{\partial}_A g g^{-1} ]
\end{equation*}
Note this equation generalizes the chirality condition
$\partial( \overline{\partial} g g^{-1} ) = 0$ that appears in
ordinary (non-fibered) WZW models.  Here, it also plays the role
of a second-class constraint.
Also note this is the supersymmetrization of the chiral anomaly in the current:
defining $j = \overline{\partial}_A g g^{-1}$, and omitting fermions,
this says $D j \propto F$.  Since we have bosonized, the anomaly is
realized classically.
In a fermionic description of the left-movers,
the current $\overline{\partial}_A g g^{-1}$ would be given by $\lambda_-
\overline{\lambda}_-$, the $[F, \overline{\partial}_A g g^{-1}] \psi \psi$
term would be a classical contribution to the divergence of the
current, and the $F$ and $\overline{\partial}(F \psi \psi)$ terms would
arise as quantum corrections, from one-loop diagrams involving
the interactions $A \lambda \lambda$
and the four-fermi term $F \psi \psi \lambda \lambda$, respectively.
The former (bosonic) contribution to the divergence is universal, 
the latter is in principle 
regularization-dependent.

\subsection{Anomaly cancellation}
In order to make the action well-defined, recall we needed to demand
that $k \int \mbox{Tr } F^2$ and $\int \mbox{Tr }R^2$ be in the same
de Rham cohomology class.  From that fact we can 
immediately read off the form of the anomaly
cancellation condition for general levels of the fibered current algebra:
if the condition at level $1$ is that
\begin{equation*}
c_2({\cal E}) \: = \: c_2(TX)
\end{equation*}
then the condition at level $k$ is
\begin{equation*}
k c_2({\cal E}) \: = \: c_2(TX).
\end{equation*}

We have already seen several independent derivations of the anomaly
cancellation condition -- it plays several roles in making the
fibered WZW model self-consistent and supersymmetric, analogues of the
same roles in heterotic worldsheets.
Here is another quick test of this claim.
Take the heterotic $E_8 \times E_8$ string on $S^1$, and orbifold by
the action which translates halfway around the $S^1$ while simultaneously
exchanging the two $E_8$'s.  The result is a theory, again on $S^1$,
but with a single $E_8$ current algebra at level two.
We can understand anomaly cancellation in this theory by working on the
covering space, before the orbifold action.
Embed bundles ${\cal E}_1$, ${\cal E}_2$ (${\cal E}_1 \cong {\cal E}_2
\cong {\cal E}$) in each of the $E_8$'s, then for anomaly cancellation to
hold we must have
\begin{equation*}
c_2({\cal E}_1) \: + \: c_2({\cal E}_2) \: = \: c_2(TX)
\end{equation*}
but this is just the statement
\begin{equation*}
2 c_2({\cal E}) \: = \: c_2(TX)
\end{equation*}
which is precisely the prediction above for anomaly cancellation in a
level two fibered current algebra.
(Attentive readers will note that the central charge of a single
$E_8$ at level two is 15.5, not 16, and so this does not suffice
for a critical heterotic string.  However, the orbifold has
massive structure in the twisted sector that is not captured
purely by the description above, and so the central charge of
the level two $E_8$ current algebra does not suffice;
put another way, in the flat ten-dimensional space limit, the $S^1$
unravels, the orbifold is undone, and some of the massive twisted sector states
become massless, curing the naive problem with the central charge.)

We can outline another derivation of the anomaly-cancellation constraint
in the language of chiral de Rham complexes \cite{edcdra,cdrc,cdrcgerb,tan}.
In those papers, the idea was to describe the perturbative physics
of a nonlinear sigma model on a space in terms of a set of free
field theories on patches on a good cover of the target space.
Conditions such as the anomaly cancellation condition arise as
consistency conditions on triple overlaps.
(Technically, the local free field descriptions need not patch
together nicely, so one need get nothing more than a stack over
the target, in fact a special stack known as a gerbe.  The anomaly
cancellation condition arises as the condition for that stack/gerbe
to be trivial.)

Here, we can follow a similar program, except that instead of associating
free theories to patches, we associate solvable theories to patches,
which is the next best thing.
So, consider the left-moving degrees of freedom, described by a current
algebra at level $k$:
\begin{equation*}
J_F^a(z) \cdot J_F^b(z') \: \sim \:
\frac{ k \delta^{ab} }{(z - z')^2} \: + \: i \sum_c
f_{abc} \frac{J^c(z')}{z-z'} \: + \: \cdots
\end{equation*}
Let $T^a$ denote the generators of the Lie algebra, and suppose that they
are functions of the base space, $T^a = T^a(\gamma(z))$ in the
notation of \cite{edcdra,tan}.
Define
\begin{equation*}
J_F(\gamma) \: = \: \sum_a J_F^a(z) T^a(\gamma(z))
\end{equation*}
Using the expansion
\begin{equation*}
T^a(\gamma(z')) \: = \: T^a(\gamma(z)) \: + \: (z'-z) \left(
\partial_{z'} \gamma^j \right) \partial_j T^a \: + \: \cdots
\end{equation*}
it is trivial to derive that the following OPE includes the terms
\begin{equation*}
J_F(\gamma(z)) \cdot J_F(\gamma(z')) \: \sim \: \cdots \: + \:
i \sum_c f_{abc} \frac{ J^c(z') }{z-z'} T^a(\gamma) T^b(\gamma) \: + \:
k \frac{ \left( \partial_{z'} \gamma^j \right) T^a(\gamma) \partial_j
T^a(\gamma) }{z-z'} \: + \: \cdots
\end{equation*}
The equation above should be compared to \cite{tan}[eqn (5.30)], for
example.  The essential difference between the two is that the
second term above (which corresponds to the fourth term on the
right-hand side of \cite{tan}[eqn (5.30)]) has an extra factor of $k$,
the level.
That $k$-dependence in the
second term on the right-hand side is ultimately responsible for
modifying the anomaly cancellation condition from 
$[ \mbox{Tr }F^2 ] = [ \mbox{Tr }R^2 ]$ to
$k [ \mbox{Tr }F^2 ] = [ \mbox{Tr }R^2 ]$.

\subsection{Massless spectra}
Letting the currents of a Kac-Moody algebra be denoted $J^a(z)$,
for $a$ an index of the ordinary Lie algebra,
the WZW primaries $\varphi_{(r)}(w)$
are fields whose OPE's with the currents have only
simple poles \cite{gincft}[section 9.1]:
\begin{equation*}
J^a(z) \cdot \varphi_{(r)}(w) \: \sim \: \frac{ t^a_{(r)} }{z-w} 
\varphi_{(r)}(w) \: + \: \cdots
\end{equation*}
where $(r)$ denotes some representation of the ordinary Lie algebra.
In other words, the WZW primaries transform under the currents just
like ordinary representations of the ordinary Lie algebra.

When we fiber WZW models, each WZW primary will define a 
smooth vector bundle associated to the principal $G$ bundle defining
how the WZW models are fibered, since across coordinate patches the
primaries will map just as sections of such a bundle.
(In the language of chiral de Rham complexes and soluble field theories
on coordinate patches, the WZW primaries transform just like sections
of associated vector bundles when we cross from one coordinate patch
to another.)
If the theory has $(0,2)$ supersymmetry, then that $C^{\infty}$
vector bundle is a holomorphic vector bundle
(otherwise, the transition functions break the BRST symmetry in the
twisted theory).

More generally, a primary together with its descendants form
a `positive-energy representation' of a Kac-Moody algebra.
Since $[J^a_0, L_0] = 0$, the states at any given mass level will
break into irreducible representations of $G$ (as described
by the zero-mode components $J^a_0$ of the currents).
(In addition, their OPE's with the full currents will have
higher-order poles, but this is not important here.)
When fibering WZW models, each such representation will then define
a vector bundle associated to the underlying principal bundle,
and so for WZW models fibered over a base manifold $X$ the states
in the positive-energy representation can be thought of as
sections of $K(X)[[q]]$, a fact which will be important to the analysis
of elliptic genera.

Following the usual yoga,
a chiral primary in the $(0,2)$ fibered WZW model is then of the form
\begin{equation*}
f_{\overline{\imath}_1 \cdots \overline{\imath}_n} \psi^{\overline{\imath}_1}
\cdots \psi^{\overline{\imath}_n}
\end{equation*}
where the $\psi$'s are right-moving worldsheet fermions, coupling to 
the tangent bundle of the base manifold $X$,
and $f$ is a section of $V \otimes \Lambda^n TX$,
where $V$ is a vector bundle defined by an irreducible representation of
$G$ corresponding to some component of a positive-energy representation
of the Kac-Moody algebra as above.  In cases\footnote{Our fibered WZW model
construction also applies to cases in which the base space is
nonK\"ahler to zeroth order in $\alpha'$.  However, that complicates
the BRST condition, and so for present purposes we restrict to Calabi-Yau
spaces.  }
in which the base space is
a Calabi-Yau to zeroth order in $\alpha'$,
for the state to the BRST closed, $f$ will be a holomorphic section,
and in fact following the usual procedure this will realize a sheaf
cohomology group valued in $V$, {\it i.e.} $H^*(X, V)$.

Morally, the integrable (or `unitary')
representations (which define WZW primaries)
correspond to massless states, as they have the lowest-lying
$L_0$ eigenvalues (though of course that need not literally
be true in all cases).

Let us briefly consider an example.
For $SU(n)$ at level 1, the integrable representations (WZW primaries)
correspond to
antisymmetric powers of the fundamental ${\bf n}$.
The construction above predicts `massless states' counted by
$H^*(X, \Lambda^* {\cal E})$ where ${\cal E}$ is a rank $n$ vector
bundle associated to a principal $SU(n)$ bundle.
These are precisely the left-Ramond-sector states described in
\cite{dg}, for ordinary heterotic worldsheets built with
left-moving fermions, and this is a standard result.
(Because \cite{dg} are concerned with heterotic compactifications,
their $SU(n)$ is embedded in $\mathrm{Spin}(16)$ and then a 
left $\BZ_2$ orbifold is performed, so there are additional
states, in $\BZ_2$ twisted sectors.)
At higher levels there are additional integrable representations.
(In fact, the integrable representations of $SU(n)$ at any level
are classified by Young diagrams of width bounded by the level.
Thus, at level 2, the adjoint representation becomes integrable,
and so in addition to the WZW current there is a WZW primary which
transforms as the adjoint.)

In ordinary heterotic compactifications, Serre duality has the effect
of exchanging particles and antiparticles.  Let us check
that the same is true here.
For any complex reductive algebraic group $G$ and any representation
$\rho$, let ${\cal E}_{\rho}$ denote the holomorphic vector bundle
associated to $\rho$.  Then on an $n$-dimensional complex manifold $X$,
Serre duality is the statement
\begin{equation*}
H^i(X, {\cal E}_{\rho}) \: \cong \: H^{n-i}(X, {\cal E}_{\rho^*} \otimes
K_X)^*
\end{equation*}
where $\rho^*$ denotes the representation dual to $\rho$.
We have implicitly used the fact that
${\cal E}_{\rho^*} \cong {\cal E}_{\rho}^{\vee}$, an immediate consequence of
the definition of dual representation (see {\it e.g.} \cite{fh}[section 8.1]).
For example, for the group $SU(n)$, the dual of the representation
$\Lambda^i V$ is $\Lambda^{i} V^*\cong \Lambda^{n-i}V$, 
exactly as needed to reproduce the
usual form.  Thus, for Serre duality on Calabi-Yau's to respect the
spectrum, properties of fields associated to representations $\rho$
must be symmetric with respect to the dual representations $\rho^*$.
Suppose the original representation $\rho$ is integrable,
then it can be shown that\footnote{The unitarity bound is \cite{gincft}[eqn (9.30)]
\begin{equation*}
2 \frac{\psi \cdot \lambda}{\psi^2} \: \leq \: k
\end{equation*}
where $\lambda$ is the highest weight of the representation in question and 
$\psi$ is the highest weight of the adjoint representation.
The highest weight of the dual representation is $- w_0 \lambda$,
where $w_0$ is the longest Weyl group element \cite{diFranc}[eqn (13.117)]. 
(The weight $- \lambda$ is the lowest weight of the dual
representation.)  Since the Killing form
is invariant under $w_0$, {\it i.e.}, $A \cdot B = (w_0 A) \cdot (w_0 B)$,
and $w_0 \psi = - \psi$, we see that the left-hand side of the inequality
is invariant under $\lambda \mapsto - w_0 \lambda$, and so 
a representation is unitary if and only if its dual is also unitary.
We would like to thank A.~Knutson for a discussion of this matter.
} the dual representation $\rho^*$ is also integrable. 
Furthermore, the conformal weights of the states are also invariant\footnote{
For a given WZW primary (which are also Virasoro primaries),
the $L_0$ eigenvalue is \cite{diFranc}[eqn (15.87)]
\begin{equation*}
h \: = \: \frac{(\lambda, \lambda+2 \rho)}{2(k+g)}
\end{equation*}
where $k$ is the level, $g$ is the dual Coxeter number, 
and $\rho$ is the Weyl vector (half-sum of positive roots).
Recall that for a highest weight $\lambda$, the highest weight of the
dual representation is $- w_0 \lambda$, where $w_0$ is the longest Weyl
group element.  Now, $w_0 \rho  = - \rho$, it takes all the positive roots
to negatives.  Thus, using the fact that the Killing metric is
Weyl invariant, 
\begin{equation*}
(\lambda, \lambda + 2 \rho) \: = \: (- w_0 \lambda, - w_0 \lambda - 2 w_0
\rho) \: = \: (- w_0 \lambda, -w_0 \lambda + 2 \rho)
\end{equation*}
and so we see that a representation and its dual define primaries
with the same conformal 
weight.
}
under this dualization. 
Thus, Serre duality symmetrically closes
states into other states, just as one would expect.

\subsection{Physical applications}
Some interesting examples of six-dimensional gauged supergravities
exist in the literature \cite{sezgin1,sezgin2,sezgin3,sezgin4},
for which a string-theoretic interpretation does not seem to be
clear at present.  The technology of this paper may give some insight
into this question.  (The relevance of higher-level currents has
been observed previously, see {\it e.g.} \cite{dienes}, but is worth
repeating here.)

One of the six-dimensional theories in question \cite{sezgin1} has a 
gauge group $E_6 \times E_7 \times U(1)$ with massless matter
in the ${\bf 912}$ representation of $E_7$.
One basic problem with realizing this in ordinary string worldsheet
constructions is that it is not clear how to build a massless
${\bf 912}$.  If we apply a standard construction, then 
the $e_7$ algebra is built from a $so(12) \times su(2)$ subalgebra.
Under that subalgebra the ${\bf 912}$ decomposes as
\begin{equation*}
{\bf 912} \: = \: ({\bf 12}, {\bf 2}) \oplus ({\bf 32}, {\bf 3}) \oplus
({\bf 352}, {\bf 1}) \oplus ({\bf 220}, {\bf 2})
\end{equation*}
However, the standard construction can only recreate adjoints (${\bf 66}$) and
spinors (${\bf 32}$) of $\mathrm{Spin}(12)$ in massless 
states from left-moving fermions,
not a ${\bf 352}$ or ${\bf 220}$, and so it is far from clear how 
a ${\bf 912}$ could arise.

By working with current algebras at higher levels, however,
more representations become unitary.  In particular, an $E_7$ current
algebra at level greater than one could have a massless state given
by a ${\bf 912}$, which is part of what one would need to reproduce
the six-dimensional supergravity in \cite{sezgin1}.
This by itself does not suffice to give a string-theoretic
interpretation of any of the six-dimensional theories described in
\cite{sezgin1,sezgin2,sezgin3,sezgin4}, but at least is a bit
of progress towards such a goal.

\subsection{Elliptic genera}
Elliptic genera are often described as one-loop partition functions
of half-twisted heterotic theories.  Since we are describing
new heterotic worldsheet constructions, we are implicitly realizing
some elliptic genera not previously considered by physicists.

However, although the elliptic genera implied by our work have not
been realized previously by physics constructions,
they have 
been studied formally in the mathematics community, in the recent\footnote{
We should briefly speak to a potential language confusion.
Many mathematics papers on elliptic genera speak of genera 
``at level $k$.''  This does not usually refer to the level of the
current algebra to which left-moving degrees of freedom couple,
but rather refers to the modular properties of the genus.
Specifically, it means the form is modular with respect to
the ``level-$k$-principal congruence subgroup'' 
$\Gamma_0(k) \subset SL(2,\BZ)$ defined by
matrices congruent mod $k$ to the identity.
Thus, Witten's elliptic genera are often called level 1 elliptic genera,
not because the left-movers couple to a level 1 current algebra,
but rather because they have good modular properties with respect
to all of $SL(2,\BZ)$.  The elliptic genera discussed in
\cite{kliu,ando1}, by constrast, have left-moving degrees of freedom
coupling to level $k$ current algebras, just as in our heterotic
fibered WZW model construction.
} works
\cite{kliu,ando1}.  Those papers describe elliptic genera in which the
left-moving degrees of freedom couple to some $G$-current algebra at
some level $k$, fibered over the base in a fashion determined by
a fixed principal $G$ bundle, just as done in this paper.

In a little more detail, each positive energy representation,
call it $E$,
of the $G$ current algebra decomposes at each mass level into a
sum of irreducible representations of $G$, and so fibering them
over the base in a fashion determined by an underlying principal
$G$ bundle $P$ yields an element $\psi(E,P) \in K(X)[[q]]$,
where the coefficient of each power of $q$ is sum of vector bundle
associated to $P$ via the irreducible representations appearing in
$E$ at the corresponding mass level.
Each such positive energy representation consists of the descendants
of some WZW primary.  The corresponding characters in an ordinary
WZW model can be interpreted as sections of line bundles over the moduli
space of flat $G$ connections on an elliptic curve \cite{looijenga}.
Replacing the coordinates on the moduli space with Chern roots of $P$
gives the Chern character of $\psi(E,P)$.
(For example, compare the $\chi_S$ in \cite{kliu}[p. 353] to
the $P_{++}$ in \cite{schwarner3}[eqn (4.15)].)

The elliptic genera described by Witten \cite{witeggen1,witeggen2}
are described and derived in this language in \cite{kliu}.
Ordinarily we think of the left-movers' contribution to Witten's elliptic
genera in terms of boundary conditions on fermions; the precise relationship
between those boundary conditions and positive energy representations
of the left-moving current algebra is spelled out in \cite{lt}[eqn (11.102)].

For the elliptic genera of \cite{witeggen1,witeggen2}, 
demanding that the genera have good modular properties implies the
standard anomaly cancellation constraint $c_2(P) = c_2(TX)$,
see for example \cite{schwarner3,schwarner1,schwarner2,lerche1}.
For fibered level $k$ current algebras,
it is shown in detail in \cite{kliu,ando1} that demanding the genera
have good modular properties implies $k c_2(P) = c_2(TX)$, the same
anomaly cancellation constraint we have already derived multiple times
from the physics of fibered WZW models.

\subsection{The relevance of principal $LG$ bundles}
We have described how to fiber WZW models,
but we (as well as \cite{kliu,ando1}) have only discussed
how to fiber in a fashion controlled
by a principal $G$ bundle with connection.
Since the WZW models describe Kac-Moody algebras,
since we are fibering current algebras,
one might expect that one could more generally fiber according to the dictates
of a principal $LG$ bundle.

Any principal $G$ bundle induces a principal $LG$ bundle,
as there is a map $BG \rightarrow BLG$.
Indeed, we have implicitly used that fact -- the Kac-Moody algebra
determined by a WZW model fits into a principal $LG$ bundle that is
such an image of a principal $G$ bundle.
If $G$ is simply-connected then a principal $LG$ bundle over $X$ can
be thought of as a principal $G$ bundle on $X \times S^1$
\cite{sm,murray,bv}.  Given a principal $LG$ bundle so described,
we can get a principal $G$ bundle just by evaluating at a point on the
$S^1$, but these maps are not terribly invertible.   
Thus, principal $LG$ bundles are not the same as principal $G$ bundles.

In fact, there is a physical difficulty with fibering Kac-Moody algebras
using general principal $LG$ bundles that do not arise from principal
$G$ bundles.  Put briefly, a physical state condition would not be
satisfied in that more general case, and so one cannot expect to
find physical theories in which left-moving current algebras have
been fibered with more general principal $LG$ bundles.

Let us work through this in more detail.
As discussed earlier, a positive energy representation of $LG$ decomposes
into irreducible representations of $G$ at each mass level,
essentially because $[ J_0^a, L_0 ] =0$.  Thus, so long as we are fibering
with a principal $G$ bundle, instead of a principal $LG$ bundle, the $L_0$
eigenvalues of states should be well-defined across coordinate patches.
(This is also the reason why the descendants can all be understood
in terms of $K(X)[[q]]$, as used in the discussion of elliptic genera.)

If we had a principal $LG$ bundle that was not the image of a principal
$G$ bundle, then the transition functions would necessarily mix up states
of different conformal weights, more or less by definition of $LG$ bundle.

Now, the physical states need to satisfy a condition of the form
$m_L^2 = m_R^2$, which defines a matching between conformal weights of
left- and right-moving parts.

In a large-radius limit, we can choose a basis of right-moving states
with well-defined $L_0$ eigenvalues.  For the left-movers, if the
WZW model is fibered with a principal $G$ bundle, then we can choose
a basis of left-moving states that also have well-defined $L_0$
eigenvalues, and so we can hope to satisfy the physical state condition 
above.  On the other hand, if the WZW model were to be fibered
with a principal $LG$ bundle, then we would not be able to choose
a basis of left-moving states with well-defined $L_0$ eigenvalues,
and would not be able to satisfy the physical state condition.

Thus, in a heterotic context, the only way to get states that satisfy
the physical state condition above is if the left-moving
current algebra couples to a principal $G$ bundle, and not a more
general principal $LG$ bundle.   

Note, however, that in a symmetrically fibered WZW model, of the form
discussed in section~\ref{symmfibwzw}, this argument would not apply.

\subsection{T-duality}
One natural question to ask is how heterotic T-duality works when
one has fibered a current algebra of level greater than one.

We have seen how the fibering structure of a fibered current
algebra is determined by a principal $G$ bundle and a connection
on that bundle.
In the special case of tori, when the flat connection over the
torus can be rotated into a maximal torus of $G$, it is
easy to speculate that heterotic T-duality should act on the
connection in a fashion independent of $k$.
After all, once one rotates the connection into a maximal torus,
the connection only sees a product of $U(1)$'s, and for $U(1)$'s
the level of the Kac-Moody algebra is essentially irrelevant.
Thus, if this conjecture is correct, in such cases heterotic T-duality
would proceed as usual.

However, even if this conjecture is correct, we have no
conjectures regarding how heterotic T-duality at higher levels should
act when the connection cannot be diagonalized into a maximal torus
(as can happen for flat connections on tori), or if the base space is
not a torus so that one only has a fiberwise notion of heterotic T-duality.

\section{Conclusions}
In this paper we have done three things:
\begin{itemize}
\item We argued that conventional heterotic worldsheet theories
do not suffice to describe arbitrary $E_8$ gauge fields in compactifications.
The basic issue is that the conventional construction builds each
$E_8$ using a $\mathrm{Spin}(16)/\BZ_2$ subgroup, and only data
reducible to $\mathrm{Spin}(16)/\BZ_2$ can be described, but not
all $E_8$ gauge fields are so reducible.
\item We reviewed alternative constructions of
the ten-dimensional $E_8$ algebra,
using other subgroups than $\mathrm{Spin}(16)/BZ_2$.
In examples we recalled the character decomposition of the
affine algebras (see {\it e.g.} \cite{kacsan} for earlier work),
and also described how that character decomposition is realized
physically in a heterotic partition function via orbifold twisted
sectors that correlate to $E_8$ group theory.
In addition to discussing maximal-rank subgroups, we also discussed
whether it may be possible to use non-maximal-rank subgroups such as
$G_2 \times F_4$.
\item We developed\footnote{After the original publication of this
paper it was pointed out to us that chiral fibered WZW models with
$(0,1)$ supersymmetry have been previously considered,
under the name ``lefton, righton Thirring models,''
see for example \cite{gates1,gates2,gates3,gates4,gates5}.
We believe we have pushed the notion somewhat further, by studying
anomaly cancellation, spectra, elliptic genera and so forth in
chiral fibered WZW models with $(0,2)$ supersymmetry.} 
fibered WZW models to describe these more general
$E_8$ constructions on arbitrary manifolds.  In fact, this allows us
to describe conformal field theories in which the left-movers couple to
general $G$-current algebras at arbitrary levels, a considerable generalization
of ordinary heterotic worldsheet constructions.  This also enables us to give
a physical realization of some new elliptic genera recently
studied in the mathematics
literature \cite{kliu,ando1}.
\end{itemize}

It would be interesting if the elliptic genera discussed here appeared in
any black hole entropy computations.

It would also be interesting to understand heterotic worldsheet
instanton corrections in these theories, along the lines of
\cite{sharpe02a,sharpe02b,sharpe02c,ade,kg}.  Unfortunately,
to produce the (0,2) analogues of the A and B models described in those
papers required a left-moving topological twist involving a global
$U(1)$ symmetry present because the left-moving fermions were
realizing a $U(n)$ current algebra at level 1.  In more general cases
there will not be such a global $U(1)$ symmetry, unless one adds it
in by hand.

\section{Acknowledgements}
This paper began in conversations with B.~Andreas and developed
after discussions with numerous other people.  Some discussion of
the initial issues regarding reducibility of $E_8$ bundles to
$\mathrm{Spin}(16)/\BZ_2$ bundles has appeared previously in
Oberwolfach report 53/2005, reporting on the ``Heterotic strings,
derived categories, and stacks'' miniworkshop held at
Oberwolfach on November 13-19, 2005.

We would like to thank M.~Ando, B.~Andreas, J.~Francis, S.~Hellerman,
M.~Hill, T.~Pantev, E. Witten and especially A.~Henriques,
A.~Knutson, E.~Scheidegger, and R.~Thomas
for useful conversations.

\appendix
\bigskip
\section{Group theory}\label{gpthy}
In this appendix we will derive some results on subgroups of
the Lie group $E_8$ that are used in the text.
We would like to thank A.~Knutson for explanations of the
material below.

First, let us collect in the following table affine Dynkin
diagrams for the simple Lie groups, labelled by the weights of
the highest-weight state for the adjoint representation, which
shall prove useful when determining subgroup structures:
\begin{equation*}
A_n: \: \xymatrix{
1 \ar@{-}[r] & 1  \ar@{-}[r] & \cdots \ar@{-}[r] & 1 \ar@{-}[r] & {*} 
\ar@{-}@/^3ex/[llll] } 
\end{equation*}
\begin{equation*}
B_n:  \: \xymatrix{
& & & &  & 1 \\
2 \ar@{<=}[r] & 2 \ar@{-}[r] & 2 \ar@{-}[r] & \cdots \ar@{-}[r] & 2
\ar@{-}[ur] \ar@{-}[dr] \\
& & & &  & {*} } 
\end{equation*}
\begin{equation*}
C_n:  \: \xymatrix{
1 \ar@{=>}[r] & 2 \ar@{-}[r] & 2 \ar@{-}[r] & \cdots \ar@{-}[r] &
2 \ar@{-}[r] & 2 \ar@{<=}[r] & {*} } 
\end{equation*}
\begin{equation*}
D_n:  \: \xymatrix{ 
1 \ar@{-}[dr] & & & & & & 1 \\
& 2 \ar@{-}[r] & 2 \ar@{-}[r] & \cdots \ar@{-}[r] &
2 \ar@{-}[r] & 2 \ar@{-}[ur] \ar@{-}[dr] \\
1 \ar@{-}[ur] & & & & & & {*} } 
\end{equation*}
\begin{equation*}
G_2: \: \xymatrix{
3 \ar@3{<-}[r] & 2 \ar@{-}[r] & {*} } 
\end{equation*}
\begin{equation*}
F_4: \: \xymatrix{
2 \ar@{-}[r] & 4 \ar@{<=}[r] & 3 \ar@{-}[r] & 2 \ar@{-}[r] & {*} } 
\end{equation*}
\begin{equation*}
E_6: \: \xymatrix{
& & {*} \ar@{-}[d] \\
& & 2 \ar@{-}[d] \\
1 \ar@{-}[r] & 2 \ar@{-}[r] & 3 \ar@{-}[r] &
2 \ar@{-}[r] & 1  } 
\end{equation*}
\begin{equation*}
E_7: \: \xymatrix{
& & & 2 \ar@{-}[d] \\
1 \ar@{-}[r] & 2 \ar@{-}[r] & 3 \ar@{-}[r] & 
4 \ar@{-}[r] & 3 \ar@{-}[r] & 2 \ar@{-}[r] & {*} } 
\end{equation*}
\begin{equation*}
E_8:  \: \xymatrix{
& & 3 \ar@{-}[d] \\
2 \ar@{-}[r] & 4 \ar@{-}[r] & 6 \ar@{-}[r] &
5 \ar@{-}[r] & 4 \ar@{-}[r] & 3 \ar@{-}[r] &
2 \ar@{-}[r] & {*} } 
\end{equation*}
(Arrows point from long to short roots.)

Next, we need to compute the centers of the universal covers
of each of the groups above.  We can read this off very simply
from the diagrams above:  the order of the center is the sum
of the number of copies of $1$ appearing on each affine Dynkin
diagram, counting the extra point $*$ as $1$.
Thus, for example, $SU(n)$ ($A_{n-1}$) has center of order $n$,
$G_2$, $F_4$, and $E_8$ have center of order 1, so no center at all,
and $E_7$ has center of order $2$, hence $\BZ_2$.
The technical reason for this is as follows.
The vertices of the affine Dynkin diagram correspond to the
corners of the Weyl alcove, corresponding to conjugacy classes
of elements whose centralizer is semisimple.  (The points in the
Weyl alcove correspond to conjugacy classes in the simply-connected
compact group.)  The label on a vertex is the order of the corresponding
conjugacy class in the adjoint group.  Any central element is its
own conjugacy class, and has semisimple centralizer -- namely,
the whole group.  Its order in the adjoint group is 1.
The result follows.

Next, to read off a maximal-rank subalgebra from one of the
affine Dynkin diagrams, first omit one of the nodes,
what remains is the Dynkin diagram for a subalgebra,
generated by all the positive roots except the one omitted.
(This will not produce all maximal-rank subalgebras in general:
to do that, one will have to repeat a process of first omitting nodes
then affinizing, possibly several times.  However, only a single step
will be required for the examples in which we are primarily interested.)

To read off a maximal-rank sub{\it group} takes a little more work.
If the node we omit is labelled above with $n$, say, then
the weight lattice for the ambient Lie algebra and the
weight lattice for the subalgebra have relative index $n$.
This means that the subgroup will have center whose order is
$n$ times larger than the center of the ambient Lie group.

For example, consider the group $E_7$.  The Lie algebra $e_7$ contains
(maximal-rank) $su(8)$, obtained by omitting the $2$ node
sticking out at the top of the Dynkin diagram.
The center of the maximal-rank subgroup of $E_7$ should then be
two times larger than that of $E_7$.  We computed above that
$E_7$ has center $\BZ_2$, hence the center of the subgroup
should have order $(2)(2) = 4$.  Now, the group $SU(8)$ has
center $\BZ_8$, so to get a center of order $4$,
we must quotient by $\BZ_2$.
Thus, a maximal-rank subgroup of $E_7$ is $SU(8)/\BZ_2$.

Similarly, we can show that $E_8$ contains the subgroup
$( E_7 \times SU(2) )/ \BZ_2$.
The subalgebra $e_7 \times su(2)$ is obtained from the affine
Dynkin diagram for $e_8$ by omitting the $2$ vertex next to the
$*$.  Thus, the center of the subgroup needs to be twice as large
as the center of $E_8$, but $E_8$ has no center, so the
center of the subgroup must be $\BZ_2$.
We computed that $E_7$ has center $\BZ_2$, and it is a standard
fact that $SU(2)$ has center $\BZ_2$, from which we deduce
that the subgroup of $E_8$ is $( E_7 \times SU(2) )/\BZ_2$.

In exactly the same fashion, one can show that 
$E_8$ has the subgroup $( E_6 \times SU(3) )/\BZ_3$.
Here we omit the $3$ node next to $2$ and $*$ on the
affine Dynkin diagram for $E_8$, which means that the center
of the subgroup must be three times larger than the center
of $E_8$, hence $\BZ_3$.  One can compute that the center
of $E_6$ is order $3$, hence $\BZ_3$, and the center of
$SU(3)$ is well-known to be $\BZ_3$, so for the
subgroup to have center $\BZ_3$ it must be
$( E_6 \times SU(3) )/\BZ_3$.

It can also be shown that $E_8$ has the subgroup
$( Spin(10) \times SU(4) )/\BZ_4$, though here
we have to work a little more.
On the affine Dynkin diagram for $E_8$, we omit the $4$ node,
and since $E_8$ has no center, we see
the subgroup should have center of order $4$.  
Both $\BZ_4$ and $\BZ_2\times \BZ_2$ are abelian of
order $4$, so we have to work slightly harder to determine whether
the subgroup is $( \mathrm{Spin}(10) \times SU(4) )/\BZ_4$
or $/ \BZ_2^2$.  In this case, since $E_8$ has no center,
the simply-connected group and the adjoint group are the same,
so the labels on the Dynkin diagram contain the element orders
in the simply-connected group, not just the adjoint group as would
ordinarily be the case.  The fact that we omitted a node marked $4$
means that the subgroup should contain a central element of order 4,
not just that the subgroup's center should be of order $4$,
from which we can deduce that the subgroup in question is
$( \mathrm{Spin}(10) \times SU(4) )/\BZ_4$.

A result that is more important for this paper is the fact that
$E_8$ contains an $(SU(5)\times SU(5))/\BZ_5$ subgroup.
We get this result by removing the $5$ node on the labelled $E_8$
Dynkin diagram.  The index of the two weight lattices is then $5$,
or put another way, the subgroup sits inside $E_8$ as the centralizer
of a certain element of (adjoint) order $5$, which is then the remaining
center.  Since $SU(5) \times SU(5)$ has center $\BZ_5 \times \BZ_5$,
we see that the subgroup of $E_8$ must be $(SU(5) \times SU(5))/\BZ_5$.

Analogous reasoning tells us that the $(SU(5) \times SU(5))/\BZ_5$
subgroup of $E_8$ above cannot
be a subgroup of $\mathrm{Spin}(16)/\BZ_2$, or vice-versa.
The $\mathrm{Spin}(16)/\BZ_2$ subgroup is obtained by removing the
leftmost $2$ node above.  The centralizer of that subgroup is then
order $2$, and because $2$ and $5$ are relatively prime, no
element of $\BZ_5$ contains an element of order $2$ or vice-versa,
hence neither is a subgroup of the other.
This result is even true at the level of algebras.
If $su(5)$ were a subalgebra of $so(8)$, then $su(5) \times su(5)$ would be
a subalgebra of $so(8) \times so(8)$, itself a subalgebra of $so(16)$,
and then there might be a way for $( SU(5) \times SU(5) )/\BZ_5$ to
be a subgroup of $\mathrm{Spin}(16)/\BZ_2$.  However, $so(8)$ does not
contain the algebra $su(5)$ -- the largest subalgebra it contains is 
$su(4) \times u(1)$.

Under the $su(5)\times su(5)$ subalgebra of $e_8$, the
${\bf 248}$ (adjoint) representation decomposes as
\begin{equation*}
({\bf 24}, {\bf 1}) \oplus ({\bf 1}, {\bf 24}) \oplus
({\bf 10}, {\bf 5}) \oplus (\overline{ {\bf 10} }, \overline{ {\bf 5} }) \oplus
({\bf 5}, \overline{ {\bf 10} }) \oplus
( \overline{ {\bf 5} }, {\bf 10})
\end{equation*}
How does the $\BZ_5$ act on the representations above?
In principle, since $E_8$ contains an $( SU(5) \times SU(5) )/\BZ_5$ 
subgroup, the representations above must be representations of
$( SU(5) \times SU(5) )/\BZ_5$, and so must be invariant
under $\BZ_5$.  Suppose the first $SU(5)$ acts on the five-dimensional
vector space $V$ in the fundamental representation, and the second
acts on $W$ in the fundamental representation.
The ${\bf 10}$'s above can be understood as the second exterior power
of $V$ or $W$.  In order for each of the representations to remain
invariant under the $\BZ_5$, the $\BZ_5$ might act on basis
elements of $V$ by fifth roots of unity, and on basis elements of
$W$ by inverses of squares of fifth roots of unity.
In other words, if $g$ denotes the generator of the $\BZ_5$, then take
\begin{equation*}
\begin{array}{rl}
g: & v \: \mapsto \: \zeta v \\
 & w \: \mapsto \: \zeta^{-2} w
\end{array}
\end{equation*}
for $v \in V$, $w \in W$, $\zeta = \exp( 2 \pi i / 5)$.
Then, with this choice of $g$ action, we see that the four non-adjoint
representations of $su(5)xsu(5)$ appearing in the decomposition of the
adjoint representation of $e_8$, namely
$(\Lambda^2 V) \otimes W$, $(\Lambda^2 V^*) \otimes W^*$,
$V \otimes (\Lambda^2 W^*)$, and $V^* \otimes (\Lambda^2 W)$,
are all invariant under $\BZ_5$.

Similarly, one can show that $E_8$ has the subgroup $SU(9)/\BZ_3$.
To get the $su(9)$ subalgebra, we omit the top $3$ node on the affine
Dynkin diagram for $E_8$, so the center of the subgroup must be three
times as large as the center of $E_8$, but since $E_8$ has no center,
we see that the center of the subgroup must be $\BZ_3$.
Since $SU(9)$ has center $\BZ_9$, we see that the subgroup of
$E_8$ must be $SU(9)/\BZ_3$.

Under the $su(9)$ subalgebra of $e_8$, the ${\bf 248}$ (adjoint) representation
decomposes as
\begin{equation*}
{\bf 80} \oplus {\bf 84} \oplus \overline{ {\bf 84} }
\end{equation*}
(The ${\bf 84}$ is $\Lambda^3 V$ for $V$ a nine-dimensional vector space,
and the ${\bf 80}$ is the adjoint representation of $SU(9)$.)
To build $E_8$ from $SU(9)$, we first quotient $SU(9)$ by $\BZ_3$.
If $V$ is a nine-dimensional vector space upon which $SU(9)$ acts in the
fundamental representation, then notice it is consistent for
the $\BZ_3$ to act as 3rd roots of unity on each element of a basis
for $V$ (consistent in the sense that the representations of $su(9)$ forming
the adjoint representation of $e_8$ are invariant under such a $\BZ_3$
 -- in other words, the representations appearing above are representations
of $SU(9)/\BZ_3$ not just $SU(9)$.)

Two cases that involve more work are the 
$su(2) \times su(8)$ and $su(2) \times su(3) \times su(6)$ subalgebras
of $E_8$.  From the analysis above, 
it is straightforward to determine
that in the first case, 
the center of the subgroup should have order $4$,
so the subgroup should have the form $SU(2) \times SU(8) / G$ for some
$G$ of order $16/4=4$, which is ambiguous.  In the second case,
the center of the subgroup should be order $6$, so the subgroup
should have the form $SU(2) \times SU(3) \times SU(6) / G$ for some
$G$ of order $(2)(3)(6)/6 = 6$, which is again ambiguous.
We can resolve the ambiguity \cite{allenpriv} 
by looking at the decomposition of the
adjoint representation of $e_8$ under each subalgebra.
In particular, in that decomposition one gets the tensor product
of fundamental representations of the factors, corresponding to
where the missing vertex is attached.  

For example,
for $su(2) \times su(8)$, the decomposition of the adjoint representation
of $e_8$ includes the $({\bf 2}, \Lambda^2 {\bf 8})$ of $su(2)\times su(8)$,
corresponding to the diagrams
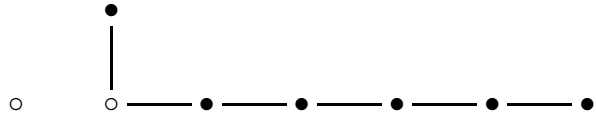
\begin{equation*}
\xymatrix{
& \bullet \ar@{-}[d] \\
\circ & \circ \ar@{-}[r] & \bullet \ar@{-}[r] & \bullet \ar@{-}[r] 
& \bullet \ar@{-}[r] 
& \bullet \ar@{-}[r] & \bullet
}
\end{equation*}
where $\circ$ indicates the position of the omitted $e_8$ diagram vertex.
Thus, inside the 
\begin{displaymath}
\BZ_2 \times
\BZ_8 \: = \: <-1> \times <t>
\end{displaymath}
center of $SU(2) \times SU(8)$, the kernel
is generated by $(-1, t^2)$.  (On the standard representation of $SU(8)$,
$t$ acts as an $8$th root of unity, but on $\Lambda^2 {\bf 8}$ it acts as
a fourth root, so $t^2$ acts as $-1$ and $(-1,t^2)$ acts as $+1$.)
Thus, the subgroup of $E_8$ with algebra $su(2) \times su(8)$ is given by
\begin{equation*}
\frac{ SU(2) \times SU(8) }{ \BZ_4 }
\end{equation*}
where the $\BZ_4$ acts diagonally.

For $su(2) \times su(3) \times su(6)$, the adjoint representation of
$e_8$ decomposes to include the $({\bf 2}, {\bf 3}, {\bf 6})$
of $su(2) \times su(3) \times su(6)$, judging by the diagrams
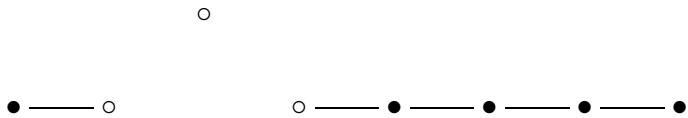
\begin{equation*}
\xymatrix{
 & & \circ \\
\bullet \ar@{-}[r] & \circ & & \circ \ar@{-}[r] & \bullet \ar@{-}[r] & 
\bullet \ar@{-}[r] & \bullet \ar@{-}[r] & \bullet
}
\end{equation*}
Inside the $\BZ_2 \times \BZ_3 \times \BZ_6 = <1> \times
<r> \times <s>$ center of $SU(2) \times SU(3) \times SU(6)$,
the kernel is generated by $(-1,1,s^3)$ and $(1,r,s^4)$,
hence the subgroup of $E_8$ with algebra $su(2) \times su(3) \times su(6)$
is given by
\begin{equation*}
\frac{ SU(2) \times SU(3) \times SU(6) }{ \BZ_2 \times \BZ_3 }
\end{equation*}
with action on the factors as indicated above.
As a consistency check, let us apply this same reasoning to the $su(5)
\times su(5)$ subalgebra of $e_8$.  Here, from omitting a vertex from
the extended Dynkin diagram, we get the diagrams
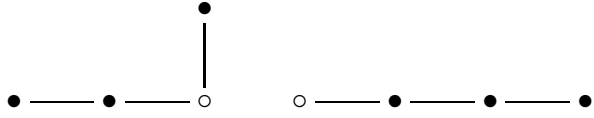
\begin{equation*}
\xymatrix{
& & \bullet \ar@{-}[d] \\
\bullet \ar@{-}[r] & \bullet \ar@{-}[r] & \circ & \circ \ar@{-}[r] &
\bullet \ar@{-}[r] & \bullet \ar@{-}[r] & \bullet
}
\end{equation*}
From the right diagram, we get a ${\bf 5}$ of $su(5)$,
and from the left diagram, we get a $\Lambda^2 {\bf 5} = {\bf 10}$.
Writing the center of $SU(5) \times SU(5)$ as $<r> \times <s>$,
the kernel is $(r^{-2}, s)$, and so the subgroup of $E_8$ with algebra
$su(5) \times su(5)$ is
\begin{equation*}
\frac{ SU(5) \times SU(5) }{ \BZ_5 }
\end{equation*}
as we worked out previously.

Since this appendix is rather lengthy, and many readers will be
most interested in simply picking off results for maximal-rank
subgroups of $E_8$, we have included a summary table below.
\begin{center}
\begin{tabular}{c}
Maximal-rank subgroups of $E_8$ \\ \hline
$(E_7 \times SU(2))/\BZ_2$ \\
$(E_6 \times SU(3))/\BZ_3$ \\
$(\mathrm{Spin}(10) \times SU(4))/\BZ_4$ \\
$(SU(5) \times SU(5))/\BZ_5$ \\
$SU(9)/\BZ_3$ \\
$(SU(2) \times SU(8))/\BZ_4$ \\
$(SU(2) \times SU(3) \times SU(6))/\BZ_2 \times \BZ_3$\\
\end{tabular}
\end{center}

Some references on these matters are \cite{borel1,borel2}.

\section{Notes on $(SU(5)\times SU(5))/\BZ_5$ bundles}
Given the role that $(SU(5)\times SU(5))/\BZ_5$ bundles play in
the analysis, we thought a short section reviewing properties
of such bundles would be useful.

First, any $SU(5)\times SU(5)$ bundle with connection
defines an $( SU(5) \times SU(5) )/\BZ_5$ bundle with
connection.  To get the bundle, one simply takes the image
of the transition functions of the original bundle in the coset,
and similarly, to get the connection, one takes the image of the
holonomies
of the original connection in the coset to get the holonomies of
the connection on the $( SU(5) \times SU(5) )/\BZ_5$ bundle.

However, the reverse need not be true -- not every $( SU(5)\times SU(5))
/ \BZ_5$ bundle defines an $SU(5)\times SU(5)$ bundle.

In addition to ordinary Chern-like invariants,
an $( SU(5) \times SU(5) )/\BZ_5$ bundle on a space $X$
has a characteristic
class in $H^2(X, \BZ_5)$, which characterizes the obstruction
to lifting to an $SU(5)^2$ bundle.  This class is defined as follows.
The short exact sequence
\begin{equation*}
\BZ_5 \: \longrightarrow \: SU(5)\times SU(5) \: \longrightarrow
\: \frac{ SU(5)\times SU(5) }{ \BZ_5 } 
\end{equation*}
extends to the right as
\begin{equation*}
\BZ_5 \: \rightarrow \: SU(5)^2 \: \rightarrow \: (SU(5)^2)/
\BZ_5 \: \rightarrow \:
B \BZ_5 \: \rightarrow \: B SU(5)^2 \: \rightarrow \:
B ( SU(5)^2 )/\BZ_5 \: \rightarrow \: K(\BZ_5, 2)
\end{equation*}

The characteristic class in $H^2(X, \BZ_5)$ comes from composing
the classifying map $X \rightarrow B( SU(5)^2 )/\BZ_5$
defining the bundle, with the map
$B (SU(5)^2)/\BZ_5 \rightarrow K(\BZ_5,2)$.

\section{$SU(N)_{1}$ characters}
The character for the ${\bf 1}$ representation of $SU(N)_{1}$ is
\begin{equation}\label{suNchar}
   \begin{aligned}
   \chi_{SU(N)}({\bf 1},\tau) = & \frac{1}{\eta(\tau)^{N-1}} \sum_{\vec{m}\in \mathbb{Z}^{N-1}} q^{(\sum m_{i}^{2} + (\sum m_{i})^{2})/2}\\
     = & \frac{1}{\eta(\tau)^{N-1}} \sum_{\vec{m}\in \mathbb{Z}^{N-1}} q ^{(\vec{m}\cdot M \cdot \vec{m})/2}
   \end{aligned}
\end{equation}
where
\begin{equation*}
   M = \begin{pmatrix}
         2 & 1 & 1 & \cdots & 1 \\
         1 & 2 & 1 & \cdots & 1 \\
         1 & 1 & 2 & \cdots & 1 \\
         \vdots&\vdots&\vdots&  & \vdots\\
         1 & 1 & 1 & \cdots & 2
       \end{pmatrix}, \qquad \text{and}\qquad
M^{-1}= \frac{1}{N}\begin{pmatrix}
           N-1& -1& -1& \cdots& -1\\
            -1&N-1& -1& \cdots& -1\\
            -1& -1&N-1& \cdots& -1\\
        \vdots&\vdots&\vdots& &\vdots\\
            -1& -1& -1& \cdots&N-1
        \end{pmatrix}
\end{equation*}
Under a modular transformation, $\chi_{SU(N)}({\bf 1},-1/\tau)$ is a 
linear combination of $\chi_{SU(N)}(\wedge^{k}{\bf N},\tau)$. 
Poisson-resumming \eqref{suNchar}, we obtain
\begin{equation}
   \begin{aligned}
   \chi_{SU(N)}(\wedge^{k}{\bf N},\tau) = & \frac{1}{\eta(\tau)^{N-1}}\sum_{\substack{\vec{m}\in \mathbb{Z}^{N-1}\\ \sum m_{i} = k\mod N}} q^{(\vec{m}\cdot M^{-1}\cdot\vec{m})/2}\\
   = & \frac{1}{\eta(\tau)^{N-1}}\sum_{\substack{\vec{m}\in \mathbb{Z}^{N-1}\\ \sum m_{i} = k\mod N}} q^{(\sum m_{i}^{2}-\tfrac{1}{N} (\sum m_{i})^{2})/2}
   \end{aligned}
\end{equation}


\begin{thebibliography}{199}

\addcontentsline{toc}{section}{References}


\bibitem{bankstalk} T. Banks, talk given at ``String vacuum workshop''
at Max-Planck Institute, Munich, November 2004, and
elsewhere.

\bibitem{vafaswamp} C. Vafa, ``The string landscape and the swampland,''
{\tt hep-th/0509212}.

\bibitem{kacsan} V. Kac, M. N. Sanielevici,
``Decompositions of representations of exceptional affine algebras with
respect to conformal subalgebras,'' Phys. Rev. {\bf D37} (1988) 2231-2237.

\bibitem{gates1} S. J. Gates, Jr., W. Siegel,
``Leftons, rightons, nonlinear sigma models, and superstrings,''
Phys. Lett. {\bf B206} (1988) 631-638.

\bibitem{gates2} D. Depireux, S. J. Gates, Jr., Q-H. Park,
``Lefton-righton formulation of massless Thirring models,''
Phys. Lett {\bf B224} (1989) 364-372.

\bibitem{gates3} S. Bellucci, D. Depireux, S. J. Gates, Jr.,
``$(1,0)$ Thirring models and the coupling of spin-zero fields to the
heterotic string,'' Phys. Lett {\bf B232} (1989) 67-74.

\bibitem{gates4} S. J. Gates, Jr., S. Ketov, S. Kuzenko, O. Soloviev,
``Lagrangian chiral coset construction of heterotic string theories
in $(1,0)$ superspace,'' Nucl. Phys. {\bf B362} (1991) 199-231.

\bibitem{gates5} S. J. Gates, Jr., ``Strings, superstrings, and
two-dimensional lagrangian field theory,'' pp 140-184 in
{\it Functional integration, geometry, and strings}, proceedings of the
XXV Winter School of Theoretical Physics, Karpacz, Poland (Feb. 1989),
ed. Z. Haba, J. Sobczyk, Birkhauser, 1989.

\bibitem{lewellen} D. Lewellen, ``Embedding higher-level Kac-Moody 
algebras in heterotic string models,''
Nucl. Phys. {\bf B337} (1990) 61-86.

\bibitem{dienes} K. Dienes, J. March-Russell, ``Realizing higher-level
gauge symmetries in string theory:  new embeddings for string GUT's,''
{\tt hep-th/9604112}.

\bibitem{bryantpriv} R. Bryant, private communication.

\bibitem{adams} J. F. Adams, {\it Lectures on Exceptional Lie Groups},
University of Chicago Press, 1996.

\bibitem{hatcher} A. Hatcher, {\it Algebraic Topology},
Cambridge University Press, 2002.

\bibitem{francispriv} J. Francis, A. Henriques, private communication.

\bibitem{dmw} E. Diaconescu, G. Moore, and E. Witten,
``$E_8$ gauge theory, and a derivation of K theory from M theory,''
Adv. Theor. Math. Phys. {\bf 6} (2003) 1031-1134, {\tt hep-th/0005090}.

\bibitem{edsymp} E. Witten, ``Global anomalies in string theory,''
pp. 61-99 in {\it Symposium on Anomalies, Geometry, Topology},
ed. by W. Bardeen and A. White, World Scientific, 1985.

\bibitem{edcmp} E. Witten, ``Global gravitational anomalies,''
Comm. Math. Phys. {\bf 100} (1985) 197-229.

\bibitem{hopkinspriv} M. Hopkins, A. Henriques, private communication.

\bibitem{edsu2} E. Witten, ``An $SU(2)$ anomaly,''  Phys. Lett. {\bf B117}
(1982) 324-328. 

\bibitem{rthompriv} R. Thomas, private communication, March 23, 2006.

\bibitem{kcs} E. Sharpe, ``K\"ahler cone substructure,''
Adv. Theor. Math. Phys. {\bf 2} (1999) 1441-1462,
{\tt hep-th/9810064}.

\bibitem{erler} J. Erler, ``Anomaly cancellation in six dimensions,''
J. Math. Phys. {\bf 35} (1994) 1819-1833.

\bibitem{fmw} R. Friedman, J. Morgan, and E. Witten, ``Vector bundles
and F theory,'' Comm. Math. Phys. {\bf 187} (1997) 679-743,
{\tt hep-th/9701162}.

\bibitem{bjps} M. Bershadsky, A. Johansen, T. Pantev, and V. Sadov,
``On four-dimensional compactifications of F-theory,''
Nucl. Phys. {\bf B505} (1997) 165-201,
{\tt hep-th/9701165}.

\bibitem{bjorn1} B. Andreas, ``On vector bundles and chiral matter in
$N=1$ heterotic compactifications,''
JHEP 9901 (1999) 011,
{\tt hep-th/9802202}.

\bibitem{bjorn2} B. Andreas, D. Hernandez Ruiperez,
``$U(n)$ vector bundles on Calabi-Yau threefolds
for string theory compactifications,'' Adv. Theor. Math. Phys.
{\bf 9} (2005) 253-284, {\tt hep-th/0410170}.

\bibitem{bjorn3} B. Andreas, G. Curio, ``Stable bundle extensions on
elliptic Calabi-Yau threefolds,'' {\tt math.AG/0611762}.

\bibitem{bjorn4} B. Andreas, G. Curio, ``Heterotic models without
fivebranes,'' {\tt hep-th/0611309}.

\bibitem{ovrut1} V. Braun, Y-H He, B. Ovrut, ``Yukawa couplings in
heterotic standard models,''
JHEP 0604 (2006) 019, {\tt hep-th/0601204}.

\bibitem{ovrut2} V. Braun, Y-H He, B. Ovrut, ``Stability of the minimal
heterotic standard model bundle,''
JHEP 0606 (2006) 032, {\tt hep-th/0602073}.

\bibitem{donagipriv} R. Donagi, private communication.

\bibitem{kollar} V. Balaji, J. Kollar, ``Holonomy groups of stable vector
bundles,'' {\tt math.AG/0601120}.

\bibitem{paulrec} P. Aspinwall, ``An analysis of fluxes by duality,''
{\tt hep-th/0504036}.

\bibitem{gswv1} M. Green. J. Schwarz, E. Witten, {\it Superstring Theory},
volume 1, Cambridge University Press, 1987.

\bibitem{gannonlam1} T. Gannon, C. Lam, ``Lattice and $\theta$-function
identities, I:  theta constants,'' J. Math. Phys. {\bf 33} (1992) 854-870.

\bibitem{diFranc} P. di Francesco, P. Mathieu, D. S\'en\'echal, {\it
Conformal Field Theory}, Springer, 1997.

\bibitem{ps} M. Peskin, D. Schroeder, {\it An Introduction to Quantum Field
Theory}, Addison-Wesley, 1995.

\bibitem{slansky} R. Slansky, ``Group theory for unified model
building,'' Phys. Rep. {\bf 79} (1981) 1-128.

\bibitem{lerchelattice} W. Lerche, A. Schellekens, N. Warner,
``Lattices and strings,'' Physics Reports {\bf 177} (1989) 1-140.

\bibitem{go} P. Goddard and D. Olive, ``Algebras, lattices, and strings,''
pp. 51-96 in {\it Vertex Operators in Mathematics and Physics},
ed. J. Lepowsky, S. Mandelstam, and I. M. Singer, Springer-Verlag, 1985.

\bibitem{cliffwzw1} P. Berglund, C. Johnson, S. Kachru, P. Zaugg,
``Heterotic coset models and (0,2) string vacua,''
Nucl. Phys. {\bf B460} (1996) 252-298,
{\tt hep-th/9509170}.

\bibitem{wittenholfac} E. Witten, ``On holomorphic factorization of WZW
and coset models,'' Comm. Math. Phys. {\bf 144} (1992) 191-245.

\bibitem{tonypriv} T. Pantev, private communication.

\bibitem{ralph1} R. Blumenhagen, A. Wisskirchen, ``Exactly solvable
(0,2) supersymmetric string vacua with GUT gauge groups,''
Nucl. Phys. {\bf B454} (1995) 561-586, {\tt hep-th/9506104}.

\bibitem{strominger} A. Strominger, ``Superstrings with torsion,''
Nucl. Phys. {\bf B274} (1986) 253-284.

\bibitem{sen1} A. Sen, ``Local gauge and Lorentz invariance of the
heterotic string theory,'' Phys. Lett. {\bf B166} (1986) 300-304.

\bibitem{sen2} A. Sen, ``Superspace analysis of local Lorentz and
gauge anomalies in the heterotic string theory,''
Phys. Lett. {\bf B174} (1986) 277-279.

\bibitem{wangwu} Z. Wang, Y.-S. Wu, ``Absence of (1,0) supersymmetry
anomaly in world-sheet gauge theories:  a purely cohomological
proof,'' Phys. Rev. {\bf D39} (1989) 509-513.

\bibitem{ssw} M. Sanielevici, G. Semenoff, Y.-S. Wu, ``Path integral
analysis of chiral bosonization,'' Nucl. Phys. {\bf B312} (1989) 197-226.

\bibitem{yuwu1} B. McClain, Y.-S. Wu, and F. Yu, ``Covariant quantization
of chiral bosons and OSP(1,1|2) symmetry,'' Nucl. Phys.
{\bf B343} (1990) 689-704.

\bibitem{yuwu2} F. Yu, Y.-S. Wu, ``Graded GL(1|1) BRST symmetry in covariant
quantization of supersymmetric chiral bosons and heterotic strings,''
Int. J. Mod. Phys. {\bf A 7} (1992) 2265-2284.

\bibitem{edcdra} E. Witten, ``Two-dimensional models with $(0,2)$
supersymmetry:  perturbative aspects,''
{\tt hep-th/0504078}.

\bibitem{cdrc} F. Malikov, V. Schechtman, and A. Vaintrob, ``Chiral
de Rham complex,'' {\tt math.AG/9803041}, {\tt math.AG/9901065},
{\tt math.AG/9905008}.

\bibitem{cdrcgerb} V. Gorbounov, F. Malikov, and V. Schechtman,
``Gerbes of chiral differential operators,''
{\tt math.AG/9906117}, {\tt math.AG/0003170}, {\tt math.AG/0005201}.

\bibitem{tan}  M.-C. Tan, ``Two-dimensional twisted sigma models and the
theory of chiral differential operators,'' {\tt hep-th/0604179}.

\bibitem{dg} J. Distler, B. Greene, ``Aspects of (2,0) string
compactifications,'' Nucl. Phys. {\bf B304} (1988) 1-62.

\bibitem{fh} W. Fulton, J. Harris, {\it Representation Theory:  A First
Course}, Grad. Texts in Math. 129, Springer-Verlag, 1991.

\bibitem{gincft} P. Ginsparg, ``Applied conformal field theory,''
lectures given at Les Houches 1988, {\tt hep-th/9108028}.

\bibitem{sezgin1} S. Randjbar-Daemi, A. Salam, E. Sezgin, J. A. Strathdee,
``An anomaly free model in six dimensions,''
Phys. Lett. {\bf B151} (1985) 351-356.

\bibitem{sezgin2} S. D. Avramis, A. Kehagias, S. Randjbar-Daemi,
``A new anomaly-free
gauged supergravity in six dimensions,''
JHEP {\bf 0505} (2005) 057, {\tt hep-th/0504033}.

\bibitem{sezgin3} S. D. Avramis, A. Kehagias,
``A systematic search for anomaly-free
supergravities in six dimensions,'' JHEP {\bf 0510} (2005) 052,
{\tt hep-th/0508172}.

\bibitem{sezgin4} D. C. Jong, A. Kaya, E. Sezgin, ``6D dyonic string with
active hyperscalars,'' {\tt hep-th/0608034}.

\bibitem{kliu} K. Liu, ``On modular invariance and rigidity theorems,''
J. Diff. Geom. {\bf 41} (1995) 343-396.

\bibitem{ando1} M. Ando, ``The sigma orientation for analytic
circle-equivariant elliptic cohomology,''
{\tt math.AT/0201092}.

\bibitem{schwarner3} A. N. Schellekens, N. P. Warner,
``Anomalies, characters, and strings,''
Nucl. Phys. {\bf B287} (1987) 317-361.

\bibitem{looijenga} E. Looijenga, ``Root systems and elliptic curves,''
Inv. Math. {\bf 38} (1976) 17-32.

\bibitem{witeggen1} E. Witten, ``Elliptic genera and quantum field
theory,'' Comm. Math. Phys. {\bf 109} (1987) 525-536.

\bibitem{witeggen2} E. Witten, ``The index of the Dirac operator in
loop space,'' pp. 161-181 in {\it Elliptic curves and modular forms in
algebraic topology}, ed. P. S. Landweber,
Lect. Notes in Math. 1326, Springer-Verlag, 1988.

\bibitem{lt} D. L\"ust, S. Theisen, {\it Lectures on String Theory},
Lect. Notes in Physics 346, Springer-Verlag, 1989.

\bibitem{schwarner1} A. N. Schellekens, N. P. Warner,
``Anomalies and modular invariance in string theory,''
Phys. Lett. {\bf B177} (1986) 317-323.

\bibitem{schwarner2} A. N. Schellekens, N. P. Warner,
``Anomaly cancellation and self-dual lattices,''
Phys. Lett. {\bf B181} (1986) 339-343.

\bibitem{lerche1} W. Lerche, ``Elliptic index and superstring
effective actions,'' Nucl. Phys. {\bf B308} (1988) 102-126.

\bibitem{sm} M. Murray, D. Stevenson, ``Higgs fields, bundle gerbes
and string structures,'' {\tt math.DG/0106179}.

\bibitem{murray} H. Garland, M. Murray, ``Kac-Moody monopoles and
periodic instantons,'' Comm. Math. Phys. {\bf 120} (1988) 335-351.

\bibitem{bv} A. Bergman, U. Varadarajan,
``Loop groups, Kaluza-Klein reduction and M-theory,''
{\tt hep-th/0406218}.


\bibitem{sharpe02a} S. Katz, E. Sharpe, ``Notes on certain (0,2)
correlation functions,'' Comm. Math. Phys. {\bf 262} (2006) 611-644,
{\tt hep-th/0406226}.

\bibitem{sharpe02b} E. Sharpe, ``Notes on correlation functions in
(0,2) theories,'' {\tt hep-th/0502064}.

\bibitem{sharpe02c} E. Sharpe, ``Notes on certain other (0,2) correlation
functions,'' {\tt hep-th/0605005}.

\bibitem{ade} A. Adams, J. Distler, M. Ernebjerg, ``Topological
heterotic rings,'' {\tt hep-th/0506263}.

\bibitem{kg} J. Guffin, S. Katz, work in progress.

\bibitem{allenpriv} A. Knutson, private communication, October 12, 2006.

\bibitem{borel1} A. Borel, J. de Siebenthal,
``Les sous-groupes fermes de rang maximum des groupes de Lie clos,''
Comment. Math. Helv. {\bf 23} (1949) 200-221.

\bibitem{borel2} A. Borel, J. de Siebenthal,
``Sur les sous-groupes fermes connexes de rang maximum des groupes de Lie
clos,''
C. R. Acad. Sci. Paris {\bf 226} (1948) 1662-1664.



\end{thebibliography}
\end{document}